\documentclass[12pt]{article}

\newcommand{\textfontsize}{11pt}
\newcommand{\sectionfontsize}{\large}
\newcommand{\subsectionfontsize}{\normalsize}
\newcommand{\marginwidth}{0.7in}
\newcommand{\titleskip}{4em}

\usepackage{graphicx}
\usepackage{setspace}
\usepackage{booktabs}
\usepackage{amsmath}

\usepackage{tabularx}
\usepackage{multirow}
\usepackage{array}
\usepackage{geometry}
\usepackage{threeparttable}
\usepackage{amssymb}
\usepackage{caption}

\usepackage[numbers, square]{natbib}

\usepackage{tocloft}
\usepackage{authblk}
\usepackage{hyperref}
\usepackage{fancyhdr}
\usepackage{titlesec}

\hypersetup{
  colorlinks=true,    
  urlcolor=blue,      
  linkcolor=blue,     
  citecolor=blue      
}

\fancypagestyle{titlepagestyle}{
    \fancyhf{} 
    \fancyfoot[R]{Page \thepage} 
}

\renewcommand{\normalsize}{\fontsize{\textfontsize}{\textfontsize}\selectfont}
\normalsize

\titleformat{\section}{\sectionfontsize\bfseries}{\thesection}{1em}{}
\titleformat{\subsection}{\subsectionfontsize\bfseries}{\thesubsection}{1em}{}

\begin{document}

\thispagestyle{titlepagestyle}

\onecolumn
\begin{center}
    \vspace*{4cm} 

    {\Large \textbf{Evaluating the Predictive Capacity of FLARES Simulations for High Redshift “Little Red Dots”}}\\[\titleskip]

    \vspace{2cm} 

    Louis Arts\textsuperscript{1\footnotemark[1]}\\
    Supervised by Jonathan Pritchard\textsuperscript{1\footnotemark[2]}\\

    \vspace{1cm} 
    \textit{Version \today}

    \vspace{2cm} 
    \includegraphics[width=0.3\textwidth]{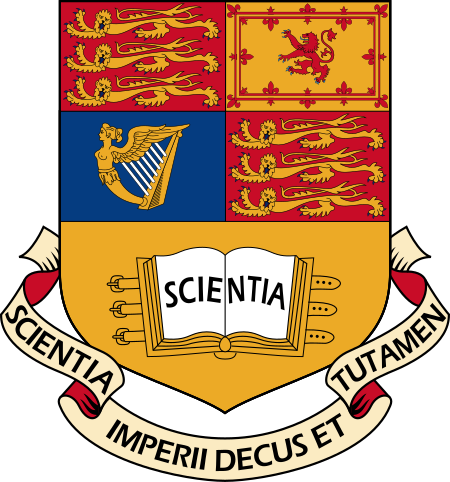} 
    \label{fig:imperial-crest}

    \vspace{2cm} 
    A thesis submitted in fulfilment of the requirements for the degree of MSc in Physics and the Diploma of Imperial College London

    \vspace{1cm} 
\end{center}

\footnotetext[1]{\href{mailto:la423@ic.ac.uk}{la423@ic.ac.uk}}
\footnotetext[2]{\href{mailto:j.pritchard@imperial.ac.uk}{j.pritchard@imperial.ac.uk}}
\footnotetext[3]{\textsuperscript{1}{\footnotesize 
\textit{Department of Physics, Imperial College London, Prince Consort Rd, South Kensington, London SW7 2BW}}}

\makeatletter
\renewcommand\@makefnmark{}
\makeatother
\newpage

\newcommand{\abstracttopskip}{10cm}

\begin{center}
    \vspace*{\abstracttopskip} 

    {\large \textbf{Declaration of Work}} 

    \begin{quote}
        \normalsize
        I hereby declare that the work presented in this thesis is my own. I was personally responsible for obtaining, processing, and analyzing the simulation and observation data used in this work. This includes the entire analysis, the development of Python code for data processing, statistical analysis, visualization, interpretation of results, and the authoring of this thesis.
    \end{quote}
\end{center}

\newpage

\begin{center}
    \vspace*{\abstracttopskip} 
    {\large \textbf{Acknowledgements}} 
    \begin{quote}
        \normalsize

        I hereby wish to thank brother John, my mother and father for their love and continued support in this endeavour. I also wish to thank my supervisor, Dr. Jonathan Pritchard, for his guidance, detailed comments, and suggestions that significantly improved this thesis. I wish to thank the FLARES team for their efforts in developing the simulation code, particularly for making their simulation output publicly available and easily accessible for use in this thesis. I want to thank Akins et al. \cite{akins2024cosmosweboverabundancephysicalnature}, the authors of the paper investigating "little red dots" through the COSMOS-Web survey, for making their processed LRD data publicly available for use in this thesis. I also wish to acknowledge the following open-source software packages used in this study: Astropy \cite{Astropy_Collaboration_and_Price-Whelan_The_Astropy_Project_2022}, HMFCalc \cite{murray2013hmfcalconlinetoolcalculating}, Scipy \cite{2020SciPy-NMeth}, Numpy \cite{2020NumPy-Array} and Matplotlib \cite{Hunter_Matplotlib_A_2D_2007}.
    \end{quote}
\end{center}

\newpage

\begin{center}

    \vspace*{13em} 
    
    {\large \textbf{Abstract}} 
    \begin{quote}
        \normalsize
        The recent discovery of "little red dots" (LRDs) - a population of extremely compact and highly dust-reddened high-redshift galaxies - by the James Webb Space Telescope presents a new challenge to the fields of astrophysics and cosmology. Their remarkably high luminosities at redshifts $5<z<10$, appear to challenge $\Lambda$CDM cosmology and galaxy formation models, as they imply stellar masses and star formation rates that exceed the upper limits set by these models. LRDs are currently subjects of debate as the mechanisms behind their high luminosities are not yet fully understood. LRD energy outputs are thought to be either dominated by star formation (starburst galaxies) or their energy output results from the hosting of active galactic nuclei (AGN). We investigate the starburst hypothesis by attempting to replicate the stellar properties of LRDs using output data from the FLARES (First Light And Reionization Epoch Simulations) simulation suite. Comparative analysis of galactic properties such as galactic number density, stellar mass and star formation rate yield significant tension between simulated and observed galaxies. The FLARES simulation overestimates the number densities of galaxies with stellar masses similar to observed LRDs by several orders of magnitude. Additionally, the simulation shows an overestimation of star formation rates. These tensions suggest a potential underestimation by the FLARES model of stellar feedback mechanisms such as active galactic nuclei feedback. These results suggest that the starburst hypothesis may be insufficient to explain the observed properties of these galaxies. Instead, the AGN scenario should be further investigated by repeating the methods in this study with a hydrodynamic galaxy simulation suite that models a higher influence of AGN feedback mechanisms on stellar activity in high-redshift galaxies.
        
        \textbf{Keywords}: little red dots, FLARES simulation, James Webb Space Telescope, active galactic nuclei, hydrodynamic galaxy simulations.
    \end{quote}
\end{center}
\newpage

\begin{spacing}{1.5} 
\tableofcontents
\end{spacing}

\newpage

\section{Introduction}
\subsection{The High Redshift Galaxy Problem}
\label{subsec:high_redshift_problem}

Less than a month after the James Webb Space Telescope recorded its first light in 2022, Labbé et al. \cite{Labb__2023} reported a population of galaxy candidates in the CEERS survey at redshifts $z>7$ with unexpectedly high stellar masses and star formation rates that appear to test the upper boundaries of what is physically possible in the \(\Lambda\)CDM paradigm. This discovery was the first in an ongoing sequence of emergent galaxy candidates at ever-increasing redshifts whose stellar properties test the upper limits imposed by our current models of galaxy formation and cosmology \cite{Adams_2022, Boylan_Kolchin_2023, carnall2024jwstexcelssurveymuch,shen2024earlygalaxiesearlydark, Atek_2022,Naidu_2022}. Much like the philosophical analogy of Bertrand Russell's teapot \cite{dawkins2024god}, the extraordinary implications associated with these galaxy candidates require rigorous and thorough validation. 

Precise measurements of cosmological parameters and galactic stellar mass distributions in the local Universe allow us to establish the theoretical maximum stellar masses of these high-redshift galaxy candidates. The cosmic baryon fraction \( f_b \), defined as the ratio of the baryonic matter density parameter \( \Omega_b \) to the total matter density parameter \( \Omega_m \), provides a straightforward way of determining these upper limits with a simple relation:

\begin{equation}
    M_{\star} \equiv \epsilon f_b M_{halo}
    \label{eq:mass_relation} 
\end{equation}

where  \(M_{halo}\) is the total baryonic mass contained in the galactic halo, including dust, gas and any other non-stellar baryonic matter, and \(\epsilon \leq 1\) is the baryon-to-star conversion efficiency with a value of \(\epsilon = 1\) corresponding to total conversion. In essence, \(\epsilon\) carries all the physics associated with star formation in the galaxy within a single dimensionless quantity. As star formation is a highly complex, multi-physics process, \(\epsilon\) is impossible to determine theoretically. Therefore, it has been inferred empirically from galactic properties in low-redshift surveys. Observations of the high-redshift regime have already revealed clear areas of tension with this model. For example, the aforementioned galaxy candidates in the CEERS survey, further investigated by Michael Boylan-Kolchin \cite{Boylan_Kolchin_2023}, were determined to have conversion efficiencies as high as \(\epsilon = 0.57\), even when allowing for 1\(\sigma\) error-adjustment. This indicates an evident tension with commonly accepted maximum conversion efficiencies of \(\epsilon \approx 0.20\), based on what we observe in the local Universe \cite{Posti_2019}.

Many of these are preliminary results, as we are still in the early stages of the JWST mission at the time of writing. Their validity rests on spectrographic confirmation of observed redshift and the elimination of systematic uncertainties (e.g., zero-point corrections, treatment of artefacts) \cite{Naidu_2022}. Onboard instrumentation is still continually being calibrated to ensure the reliability of data products \cite{boyer2022jwst}. As such, the assumed redshifts of observations might be incorrect. Attributing these results to calibration errors, we could assume that the observed high stellar mass galaxies are at lower redshift than previously thought. If this is true, there would be no need to impose a high stellar conversion rate as these galaxy candidates would have had more time to form stars than previously thought. However, a range of observations has recently been spectrographically confirmed. Additionally, although not as precise as spectrographic confirmation, galaxy SED model fitting and photometric redshifts can still offer sufficient accuracy for testing our cosmological and galaxy formation models against real-world data. 


\subsection{Little Red Dots}
\begin{figure}[!htb]
  \centering
  \vspace{0mm} 
  \includegraphics[width=0.7\linewidth]{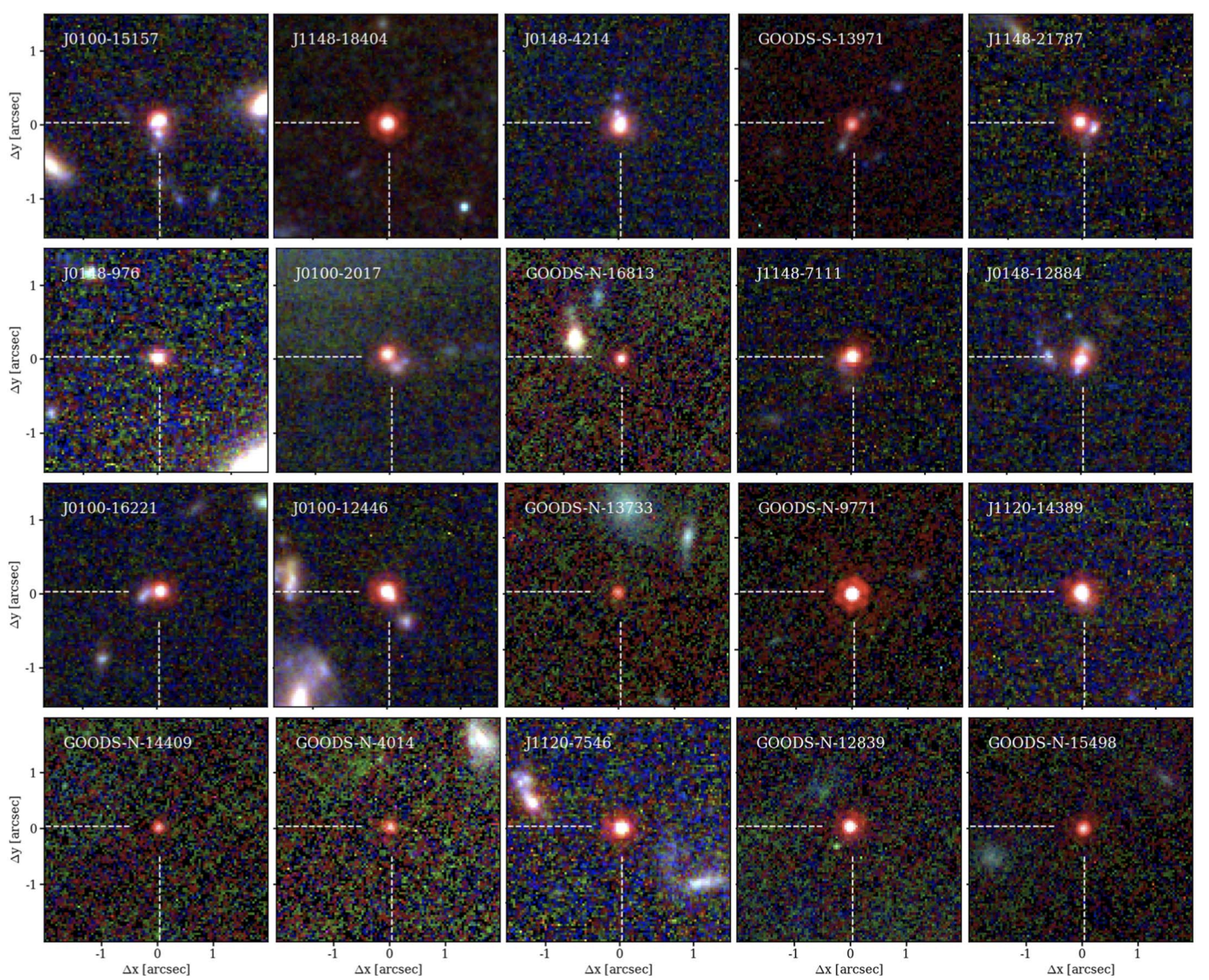} 
  \vspace{-3mm} 
  \caption[Short caption for List of Figures]{False colour stamps of 20 little red dots at $z = 4.2 - 5.5$ identified and presented by Matthee et al. (2024) \cite{matthee2024littlereddotsabundant} using NIRCam imaging of the EIGER and FRESCO surveys. Little red dots stand out as red point sources}
  \label{fig:DOTSSSS}
\end{figure}

A particularly enigmatic class of "problematic" early Universe galaxies observed by the JWST are known as "little red dots" (LRDs). These galaxy candidates were recently observed at high redshifts ($z>5$), corresponding to a time in cosmic history when the Universe was less than one billion years old. These galaxies were given their name for their small angular size and reddish appearance in JWST images. These characteristics are visualised in Figure~\ref{fig:DOTSSSS}, which shows false colour stamps of 20 little red dots at redshifts at $z = 4.2 - 5.5$ identified by Matthee et al. (2024) \cite{matthee2024littlereddotsabundant} using NIRCam imaging of the EIGER and FRESCO surveys. These galaxies appear far redder than what is expected from their cosmological redshifts. The significant red appearance is believed to be partly caused by heavy dust obscuration, which is unexpected in these young cosmic environments \cite{kocevski2024risefaintredagn}. Additionally, the effective radii of some LRDs have been measured to be as small as 10-20 pc with a median effective radius of 80 pc, significantly smaller than expected for galaxies of comparable mass \cite{Guia_2024}. 

Their distinctive characteristics and unexpectedly high abundance in various JWST surveys have sparked debate over the underlying cause of these signatures \cite{ananna2024xrayviewlittlered}. Two distinct hypotheses exist: these galaxies are either dominated by extreme star formation (starburst galaxies) or host active galactic nuclei.  The effective fit of spectral energy distribution (SED) models matching intense and compact starbursts provides the primary supporting data for the starburst scenario. These SED models reflect some of the spectral traits of LRDs, including the intense emission from young OB stars, suggesting that active star formation rather than AGN activity drives their energy output. The lack of observable X-ray emission from LRDs further undermines the hypothesis of substantial AGN activity and instead supports the scenario where star formation is the primary source of their energy output. Active galactic nuclei are known to be significant emitters of X-rays through a process known as inverse Compton scattering. Above the accretion disc of an AGN, a region known as the corona is composed of high-energy electrons which upscatter low-energy photons from the accretion disc, producing high-energy X-rays. The detection of X-rays is also considered one of the least biased indications of AGN activity in galaxies since X-rays can penetrate through a large amount of obscuring material that might otherwise block optical and UV radiation \cite{Brandt_2015}.

Conversely, the AGN hypothesis is backed by the observation of broad H$\alpha$ emission lines in certain LRDs, which strongly suggests the existence of supermassive black holes (SMBH) \cite{wasleske2024activedwarfgalaxydatabase}\cite{ananna2024xrayviewlittlered}. These broad lines would result from Doppler broadening due to high-speed
accretion of line-emitting gas near the black hole of up to $ \sim 2.5 \times 10^3 \, \text{km} \, \text{s}^{-1} $ \cite{wang2024rubiesevolvedstellarpopulations}, e.g. from the rapid receding and approaching portions of a rapidly spinning accretion disc. The compactness and point source light profiles of these LRDs additionally point to a possible AGN scenario as these features are characteristic of a highly compacted central "engine", such as an accretion disc, significantly outshining the host galaxy. Alternatively, LRD compactness has been proposed to be a consequence of a low-spin scenario. In this scenario, LRDs represent the low spin tail of the distribution of specific angular momenta (angular momentum per unit halo mass) of high-redshift galaxies. Mo et al. (1998) \cite{Mo_1998} presents a semi-analytical model relating galaxy radius to specific angular momentum. As such, the low-spin tail argument would explain the observed compactness of LRDs.

\captionsetup{font=scriptsize, width=0.7\linewidth} 

\subsection{Hydrodynamic Galaxy Simulation}
\subsubsection{Overview}

Assessing the baryon-to-star conversion efficiencies for these galaxy candidates provides an initial validation method. Although this approach is thorough, it does have its constraints. The main challenge arises from the fact that this model does not entirely reflect the intricacies of galaxy formation at high redshifts, especially in conditions very different from the local Universe. Furthermore, this approach only allows for a rudimentary evaluation by limiting the analysis to a single dimensionless metric. Although this comparison can identify possible initial tensions, it does not adequately address the broader mechanisms involved in galaxy formation. Consequently, essential metrics such as galactic number densities ($\phi / \text{Mpc}^{-3}$) and star formation rates ($\text{M}_\odot \, \text{yr}^{-1}$) do not play a role in this type of analysis. One particularly effective verification method is the comparison of real-world observations to theoretical predictions from hydrodynamic galaxy simulations. 

These computational models integrate a set of hydrodynamic differential equations forward in time starting from specific initial conditions set by measured concordance models of cosmology \cite{vogels}, which have been measured with high accuracy through surveys such as Planck \cite{PLANCK} and WMAP \cite{WMAP}. In these simulations, the selection of initial conditions is crucial. A perturbation field on a selected base $\Lambda$CDM cosmology sets up these conditions.  This background cosmology is usually specified as a flat FLRW metric with a given set of cosmological density parameters for dark matter $\Omega_{c}$, dark energy $\Omega_{\Lambda}$ and baryonic matter $\Omega_{b}$. To obtain a matter power spectrum, Gaussian density perturbations are then applied to this background density field, as predicted by inflationary theories \cite{Eisenstein_1999}.

A matter power spectrum defines the matter distribution in the simulated space at several scales. This is computed by breaking down the perturbation field into its component density scale modes using a Fourier transform. Typically, \(z \approx 100 \) is the redshift at which the density perturbation field (starting condition) is specified, and it is then integrated forward in time from that point \cite{vogels}. As such, researchers can simulate the dynamics associated with early Universe galaxy formation inside a virtual "box" at a chosen volume and resolution containing components such as dark matter, dark energy and baryonic matter at specified densities. Since the selection of initial conditions can greatly affect the simulated result, these models need to be precisely calibrated against empirical data to maintain their scientific validity \cite{Crain_2015}.

Hydrodynamic galaxy simulations essentially combine the effects of gravity and hydrodynamics. The hydrodynamic aspect of the simulation models the evolution of the simulated gas's velocity field over time. The Navier-Stokes equations are the primary equations that describe these dynamics. On the other hand, the Poisson equation for gravity links the gravitational potential to the mass distribution of the simulated system. These are the two primary sets of differential equations governing the dynamics of galaxies. However, additional dynamics, including thermodynamics and magneto-hydrodynamics, are
often introduced to obtain a more detailed and physically accurate simulation \cite{Schaye_2014}.
Furthermore, many simulations consider star formation, the Universe's expansion, and feedback mechanisms from active galactic nuclei and supernovae. 

These simulations are valuable tools in cosmology as they allow us to numerically model various complex physical processes associated with galaxy formation \cite{Schaye_2014}. New real-world observations can then be tested against the outputs of these simulations to verify whether our current models of cosmology are in tension with what we observe (or vice versa) \cite{2023MNRAS.524...43T}.

\noindent
A significant drawback to these simulations, however, is the computational expense associated with them. Hydrodynamic galaxy simulations are notoriously expensive to run as they often model multi-physics, highly non-linear processes. In addition to this, the high resolution and volume over which these dynamics need to be simulated significantly add to their computational cost \cite{vogels}. Higher-resolution simulations, which model more detailed physical phenomena such as the formation of individual galaxies or smaller galaxy populations, often require significantly higher computing power and memory. On the other hand, lower-resolution/higher-volume simulations, which are helpful for statistical inference, also become computationally expensive. As such, a trade-off exists between resolution and volume. Often, a simulation project will consist of large-volume simulations and low-volume/high-resolution "zoom-in" re-simulations of overdense regions. 

Another trade-off in computational expense is the inclusion or exclusion of some physical processes. The most prominent example of this is the inclusion or exclusion of baryonic matter in the model. Dark matter-only simulations approximate ignoring the effects of baryonic matter and only simulate the dynamics of collisionless dark matter \cite{Vogelsberger_2014} because the abundance of dark matter ($\Omega_{c} h^{2} = 0.120 \pm 0.001$) is significantly higher than baryonic matter ($\Omega_{b} h^{2} = 0.0224 \pm 0.0001$) in $\Lambda$CDM cosmology \cite{PLANCK}. This leads to a much simpler, pressureless model requiring less computing power. These simulations nevertheless offer essential insights into the formation of galaxies since dark matter halos supply the gravitational wells that catalyse the formation of large-scale structures. As such, they offer a framework upon which more intricate models incorporating baryonic physics may be built. The computational advantage of these approximations also allows for larger-scale simulations over more extended periods of cosmic time, allowing for better statistical inference. Even when excluding the impacts of baryonic gas, which primarily interacts through relatively short-range forces, structure formation can be accurately simulated on these larger scales since dark matter mainly interacts through the very long-range force of gravity \cite{vogels}. Because of these necessary trade-offs, various simulation suites have been developed, each optimised for different aspects of galaxy formation and varying widely in their methodologies, implementations and applications.

\subsubsection{The FLARES Simulation}
\label{subsec:FLARESsimul}

FLARES (First Light And Reionization Epoch Simulations) \cite{wilkins_flares_2023} is a recently developed hydrodynamic galaxy simulation project whose data has not yet been extensively compared to the recent JWST LRD observations. FLARES is a "zoom-in" simulation suite focused on galaxy formation during the Epoch of Reionisation (EoR). Based on the previously developed large-volume dark-matter only EAGLE (Evolution and Assembly of GaLaxies and their Environments) simulation model, FLARES re-simulates several smaller spherical regions from a more extensive $(3.2\, \mathrm{cGpc})^3$ parent volume within a redshift range of $z = 5 - 10$ \cite{flares}. 

\begin{figure}[!htb]
  \centering
  \vspace{0mm} 
  \includegraphics[width=0.8\linewidth]{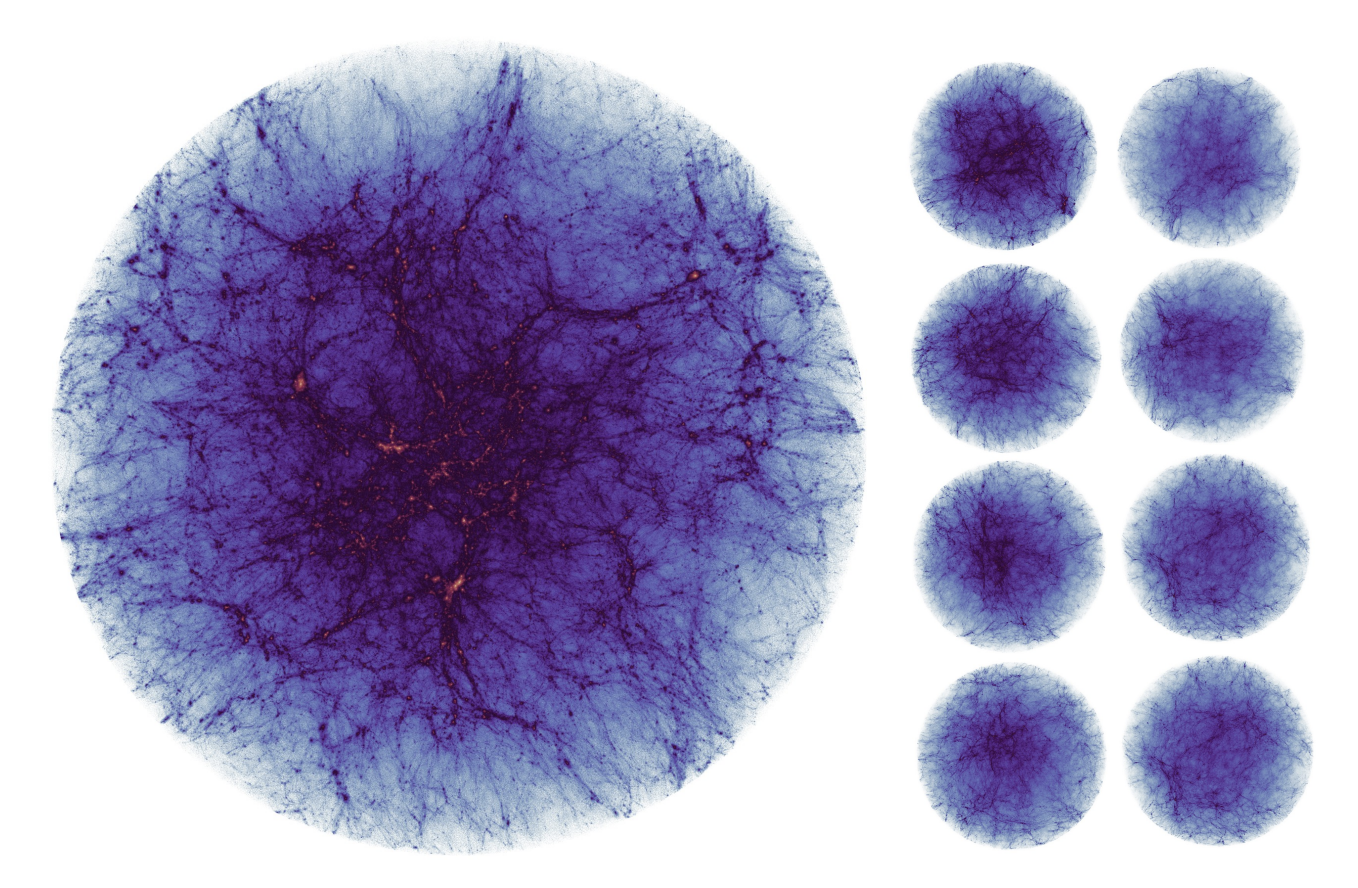} 
  \vspace{-3mm} 
  \captionsetup{width=0.8\linewidth}
  \caption[Short caption for List of Figures]{Visualisation of 9 FLARES resimulations of overdense regions of the larger 3.2 cGpc side length volume (dark matter integrated). Retrieved from a paper by Lovell et al. (2020) \cite{flares}.}
  \label{fig:BLUEVISS}
\end{figure}

A visualisation of these spherical re-simulated regions is given in Figure \ref{fig:BLUEVISS}. Based on overdensity—the relative density of matter relative to the cosmic mean—FLARES extracts particular regions from the parent simulation and simulates them with full hydrodynamics at a considerably higher resolution. This eliminates the need to re-simulate the entire Universe at high resolution and enables the researchers to obtain high-quality, high-resolution data for specific regions of interest. This then allows for the inclusion of baryonic physics, such as star formation and feedback mechanisms, in these smaller re-simulations while circumventing the high computational cost associated with including baryonic physics.

A weighing scheme is applied to construct composite distribution functions representing the larger parent volume. This approach allows for a wide dynamic range, covering galactic densities from the most overdense regions to underdense voids. FLARES then employs the Friends-of-Friends (FOF) algorithm to identify galaxies and the Subfind algorithm to split them up into gravitationally bound substructures, which are then treated as individual galaxies. This allows one to examine a single galaxy in detail and determine critical features such as its stellar mass and star formation rate. One of FLARES' key features is the ability to model feedback from stars and active galactic nuclei on the interstellar medium. This feedback plays a crucial role in regulating star formation, affecting the increase of stellar mass, and determining star formation quenching.

In hydrodynamic galaxy simulations, there is usually a robust negative correlation between the simulated volume and the achievable resolution. As shown in Figure \ref{fig:comp} (reproduced from the work of Lovell et al. \cite{flares}), the re-simulation method employed by FLARES distinguishes it from comparable simulations by significantly extending its volume reach relative to its achievable resolution. 

\captionsetup{font=scriptsize} 

\subsection{Research Aim and Significance}

The main objective of this research is to assess the predictive capability of FLARES simulations in determining the characteristics of high-redshift LRDs, assuming that the starburst hypothesis is accurate, meaning that the energy production of these galaxies is primarily influenced by star formation. The primary objective of this project is to reproduce measurements such as stellar mass, star formation rate, specific star formation rate, and galactic number densities. This will be done by comparing simulated data with observations within the same redshift range.

In 2023, three separate studies by Kannan et al. \cite{Kannan_2023}, McCaffey et al. \cite{McCaffrey_2023} and Geraldi et al. \cite{garaldi2023thesan} compared the high stellar mass JWST observations with outputs from the MilleniumTNG, Renaissance, and Thesan simulation suites respectively. Remarkably, every group found that their simulations could reproduce at least a subset of high redshift JWST galaxy observations. However, they often fail to reach the required resolution and integration of physical processes to fully shed light on the formation of these high stellar mass galaxies. Applying these comparison methods across a broader range of simulation projects is thus essential. Analysing data from diverse types of simulations against new JWST observations helps to overcome the previously mentioned lack of computing power by leveraging the strengths of different simulation approaches.

Although FLARES simulates galaxy formation in the same redshift range as observed LRDs, and its data products were designed to be compared to JWST observations, no extensive study has compared its outputs to the recent LRD observations at the time of writing. Successful reproduction of galactic properties supports the starburst scenario and confirms FLARES' effectiveness in the high-redshift regime. Alternatively, discrepancies in stellar characteristics or galactic abundances may indicate that the starburst scenario is an inadequate explanation for the observed properties of LRDs and would thus suggest the AGN hypothesis as the more likely explanation.

\section{Data}
\subsection{COSMOS-Web}

\begin{figure}[!htb]
  \centering
  \vspace{0mm} 
  \begin{minipage}[t]{0.45\linewidth} 
    \centering
    \includegraphics[width=\linewidth]{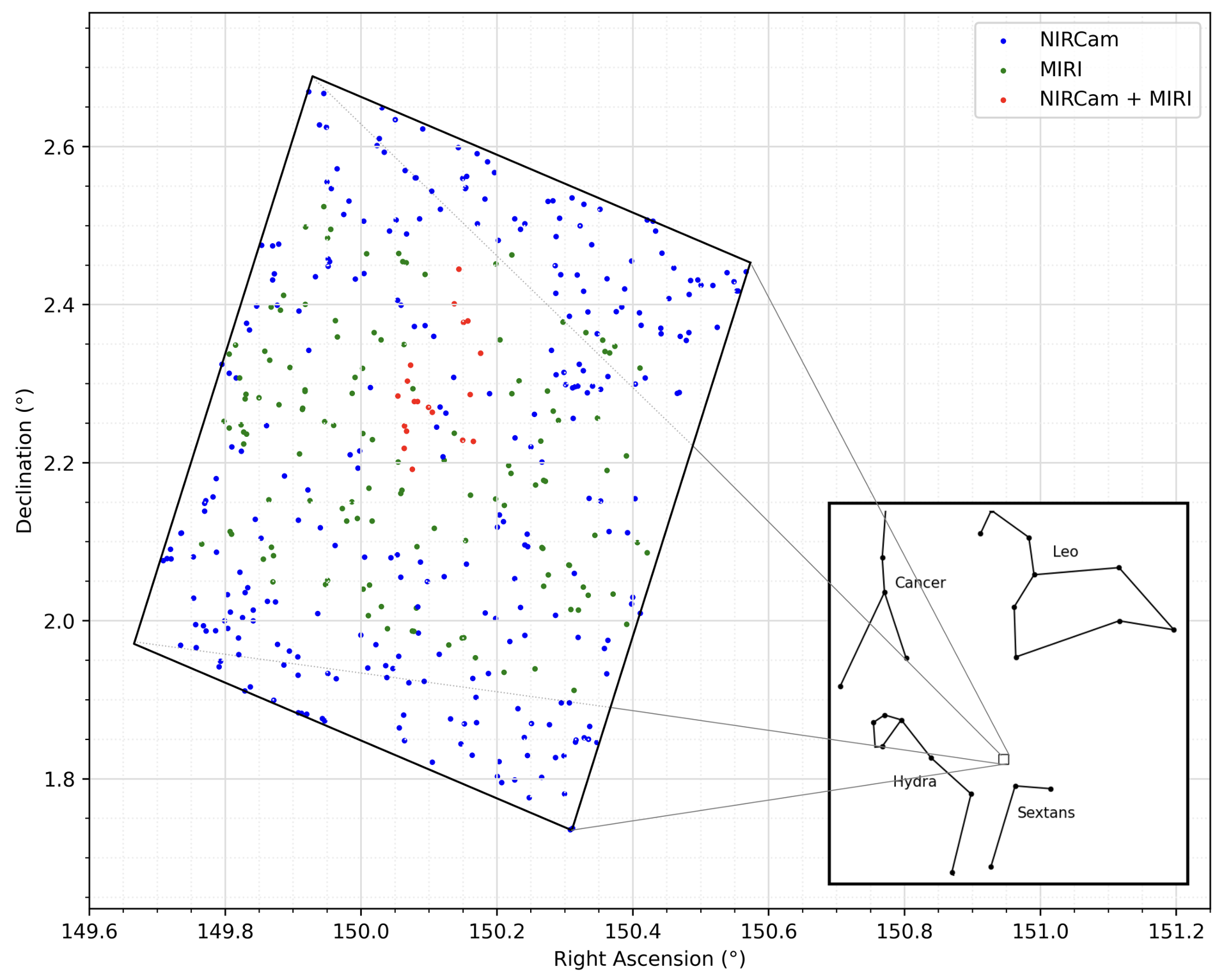}
    \captionsetup{width=1\linewidth}
    \caption[Short caption for List of Figures]{Declination vs. right ascension of COSMOS-Web galaxy candidates captured by the NIRCam (blue), MIRI (green), and both (red). The angular size of the entire survey is compared to the Cancer, Leo, Hydra, and Sextans constellations (bottom right).}
    \label{fig:surveyarea}
  \end{minipage}
  \hspace{0.05\linewidth} 
  \begin{minipage}[t]{0.45\linewidth} 
    \centering
    \includegraphics[height=7.1cm]{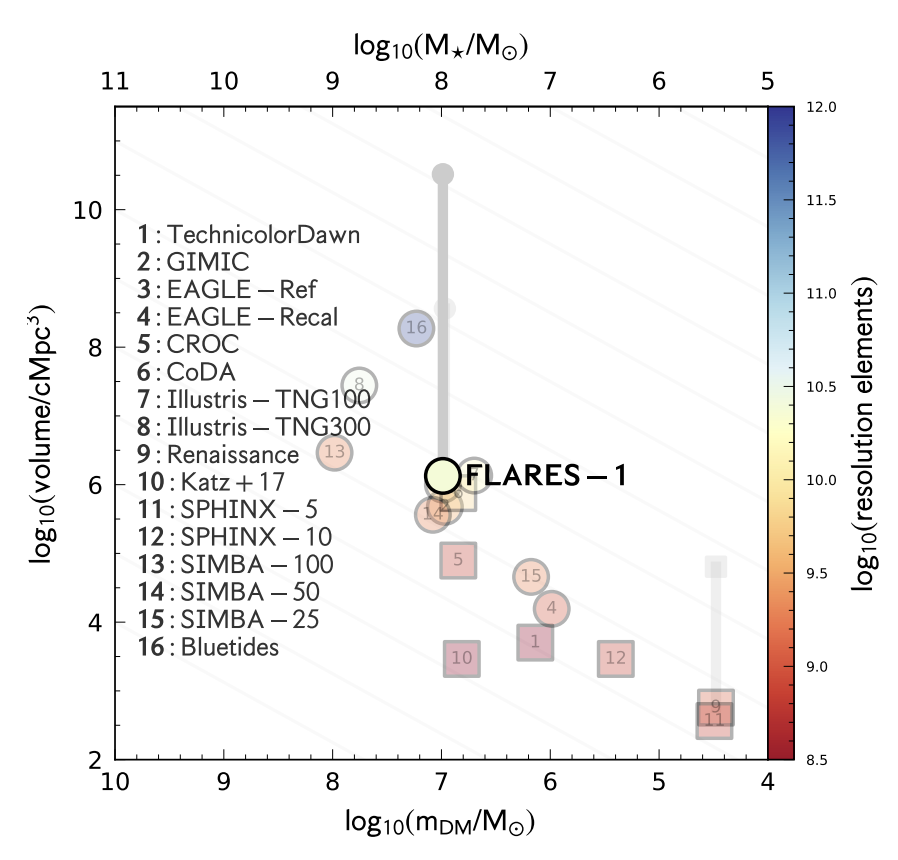} 
    \captionsetup{width=1\linewidth}
    \caption[Short caption for List of Figures]{Comparison of the resolution of dark matter elements to the simulated volume for various high redshift galaxy simulations. Compared to other simulations, FLARES demonstrates a higher resolution for its volume. The figure illustrates a robust negative correlation between the achievable resolution and the simulated volume. Generally, larger volumes are associated with lower resolution, except re-simulation methods such as FLARES, which considerably extend the volume axis. This figure is reproduced from the work of Lovell et al. \cite{flares}}
    \label{fig:comp}
  \end{minipage}
\end{figure}

The LRD data in this work is sourced from the COSMOS-Web survey, a large-scale JWST Cycle 1 treasury program that was designed to provide extensive imaging of the COSMOS field. The Mid-Infrared Instrument (MIRI) and the Near Infrared Camera (NIRCam) were employed by the JWST to survey a contiguous area of 0.19 square degrees and 0.54 square degrees, respectively. Figure \ref{fig:surveyarea} shows a visualisation of the relative area covered by both instruments. Relevant data from a total of 434 LRDs were obtained from this dataset. Raw JWST data were not processed in this study; rather, pre-processed datasets were employed, as supplied by Akins et al. (2024) \cite{akins2024cosmosweboverabundancephysicalnature}. Pre-processed LRD data of interest in this study included UV magnitudes, stellar masses, redshifts, and star formation rates.

The JWST Calibration Pipeline was employed by Akins et al. to process the raw JWST NIRCam and MIRI imaging data for the COSMOS-Web survey. The pipeline was customized to the specific instrument versions (NIRCam: version 1.12.1; MIRI: version 1.8.4). The NIRCam data underwent supplementary custom modifications, including removing 1/f noise and sky background. To guarantee a median offset in RA and Dec of less than five milliarcseconds, the final mosaics were generated using precise astrometric calibration linked to Gaia-EDR3. The MIRI data also included additional steps for background subtraction to address instrumental effects \cite{akins2024cosmosweboverabundancephysicalnature}.

 Akins et al. investigated the two extreme scenarios of LRDs by utilising starburst and AGN models to derive stellar masses and star formation rates through SED fitting. However, only starburst-fitted data products were used in this study to maintain consistency with the original research question. A combination of spectroscopic and photometric methodologies was employed to determine LRD redshifts. Photometric redshifts, derived from multi-band JWST data, provide valuable estimates; however, they are subject to inherent uncertainties due to the absence of spectroscopic coverage for the entire sample. All inherent uncertainties in galactic properties and redshifts were considered in the final analysis. An overview of relevant galactic property distributions and pairwise relationships between metrics in the COSMOS-Web dataset is given in figure \ref{fig:pairplotcosmos}. 

\captionsetup{font=scriptsize} 

\subsection{FLARES}

\begin{figure}[!htb]
  \centering
  \vspace{0mm} 
  \begin{minipage}[t]{0.45\linewidth} 
    \centering
    \includegraphics[width=\linewidth]{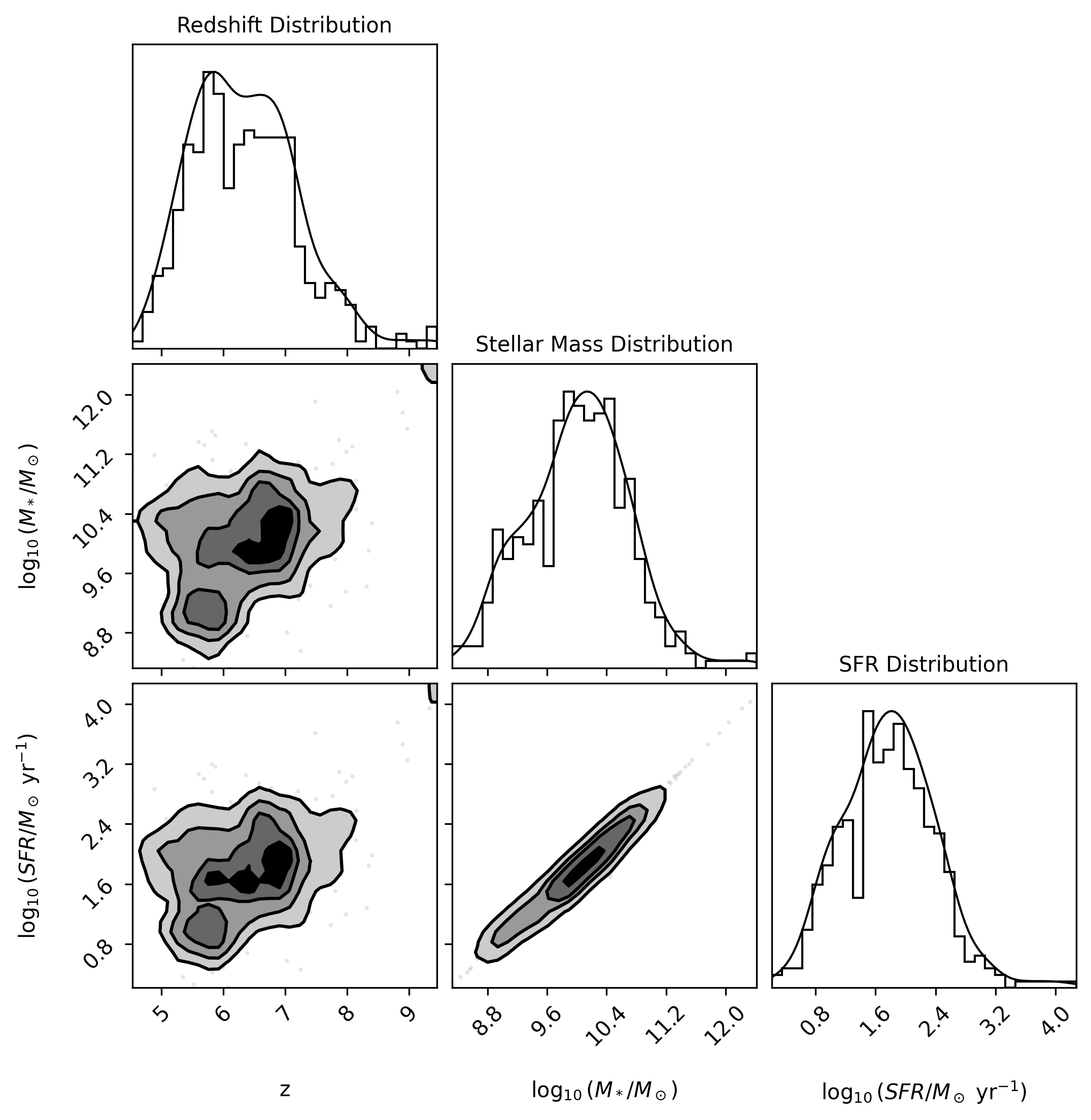}
    \vspace{0mm} 
    \captionsetup{width=1\linewidth}
    \caption[Short caption for List of Figures]{Distributions and correlations of galaxy properties from the COSMOS-Web survey. Top left: Redshift distribution with a black KDE fit line. Top middle: Stellar mass distribution with a black KDE fit line. Top right: Star formation rate (SFR) distribution with a black KDE fit line. The three panels in the bottom left corner depict contour maps illustrating the relationships between redshift, stellar mass, and SFR across the galaxy sample.}
    \label{fig:pairplotcosmos}
  \end{minipage}
  \hspace{0.05\linewidth} 
  \begin{minipage}[t]{0.45\linewidth} 
    \centering
    \includegraphics[width=\linewidth]{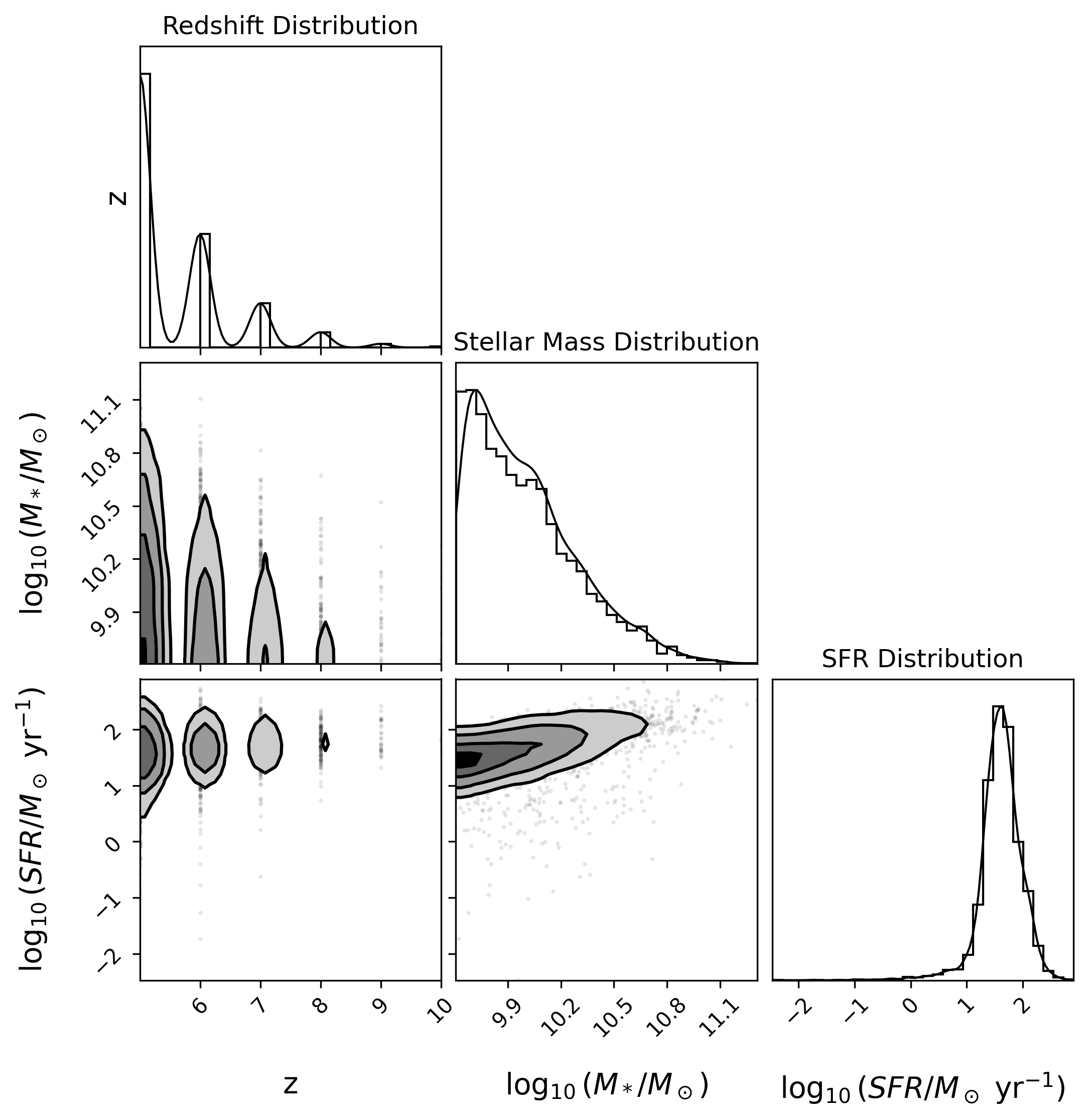}
    \vspace{0mm} 
    \captionsetup{width=1\linewidth}
    \caption[Short caption for List of Figures]{Distributions and correlations of galaxy properties from the FLARES simulated galaxies. Top left: Redshift distribution with a black KDE (Kernel Density Estimate) fit line. Top middle: Stellar mass distribution with a black KDE fit line. Top right: Star formation rate (SFR) distribution with a black KDE fit line. The three panels in the bottom left corner depict contour maps illustrating the relationships between redshift, stellar mass, and SFR across the galaxy sample.}
    \label{fig:pairplotflares}
  \end{minipage}
\end{figure}

Simulated FLARES data was obtained from the FLARES—Data Release I. This release contains data on 704742 individual simulated galaxies covering 40 re-simulated "zoom-in" regions with key snapshots at redshifts $z=5$, $z=6$, $z=7$, $z=8$, $z=9$, and $z=10$.  The parent dark matter EAGLE simulation uses a particle mass of $8.01 \times 10^{10} \, M_{\odot}$.  At the same time, the re-simulated regions achieve a much higher resolution, particularly in dense regions, with a gas particle mass of $1.8 \times 10^{6} \, M_{\odot}$ and a softening length of $2.66 \, \text{ckpc}$. Available galactic properties in the Data Release I are limited to redshift, stellar mass and star formation rate. An overview of relevant galactic property distributions and pairwise relationships between metrics in the FLARES dataset is given in figure \ref{fig:pairplotflares}. 

\captionsetup{font=scriptsize} 

\section{Methods}

\subsection{Mock Observations}
\label{subsec:mock}

Hydrodynamic galaxy simulations allow us to study galactic properties such as stellar mass, luminosities and star formation rates at cosmic times over several orders of magnitude. These are powerful tools in cosmology as they allow us to access accurate data on every galaxy we simulate. These simulations do not encounter the factors that commonly hinder our capacity to acquire low-error data in observation, such as constraints in telescope sensitivity, dust obscuration, and the inherent dimness of these remote objects. It is thus vital to consider these conditions when directly comparing simulated galaxy populations to real-world data from JWST surveys. As such, mock observations were first generated from the FLARES data. This was done by filtering out the galaxies that would not be observable by the JWST. As such, we facilitate a more direct comparison of FLARES simulations to real-world observations.

The FLARES--Data Release 1 only provide stellar masses and star formation rates for simulated galaxies. Therefore, absolute UV magnitudes were first derived from galactic stellar masses using a mass-to-light ratio derived by Grazian et al. \cite{Grazian_2015}. Specifically, the relation used is given by:

\begin{equation}
\log M_\star = -0.4 \times M_{\text{UV}} + 1.6
\end{equation}

This relation, derived from observational data of galaxies in the redshift range \(3.5 < z < 4.5\) in the GOODS-South field, has been shown to remain predictive even at higher redshifts \cite{Grazian_2015}. After converting the stellar masses from the FLARES simulations to corresponding UV magnitudes, we applied a selection criterion to match the observational capabilities of the COSMOS-Web survey. The faintest galaxy detected in COSMOS-Web survey data corresponds to an absolute UV magnitude of \(M_{\text{UV}} = -20.015\). Therefore, we filtered the simulated FLARES galaxies, excluding those with UV magnitudes higher than this threshold.

This selection process reduced the initial FLARES dataset from 704,742 galaxies to a subset of 3,542 galaxies, representing a reduction of approximately 99.5\%. This filtered subset constitutes our mock observations and is then directly compared to observation.

\subsection{Galaxy Stellar Mass Functions}

Apart from the stellar properties of individual galaxies, a critical metric compared between the COSMOS-Web survey and the FLARES simulation in this study is galactic number densities. Graphing the number density of galaxies in function of stellar mass allows us to generate galaxy stellar mass functions. They serve as an essential indicator of the characteristics of galaxy populations. Generating these functions at various redshifts allows us to follow the evolution of galaxy populations and the assembly of stellar mass across cosmic time. The FLARES simulation allows us to compute highly detailed GSMFs thanks to a high dark matter particle mass resolution of $5 \times 10^4 M_{\odot}$ and a gas particle mass resolution of $1.8 \times 10^6 M_{\odot}$ \cite{flares}. These distributions can then be fitted to facilitate a more straightforward comparison to JWST observations. 

To facilitate this, we first calculate the co-moving volumes in which the galaxies are simulated/observed. In the case of the FLARES simulation, this is a straightforward calculation. The FLARES—Data Release I consists of 40 snapshots, taken at six distinct redshifts, simulating a spherical region of radius 14 $h^{-1}$ Mpc for each redshift. We choose $h=0.674$ and all other cosmological parameters in this study based on the Planck 2018 results \cite{PLANCK}. Multiplying this volume by the number of snapshots at a given redshift gives us a total volume of 1,352,513 $\text{Mpc}^3$ at that redshift. Since galaxies were simulated at six distinct redshifts, we group them into redshift bins centered around each of these redshifts, effectively covering the entire redshift range. As such, we divide the total number of simulated galaxies at a given redshift by this volume to get a number density at this given redshift (bin). 

Similarly, we subdivided the COSMOS-Web galaxies into six redshift bins centred around the FLARES redshifts. We then calculate the co-moving observed volume for each redshift bin separately. The co-moving volume was calculated as follows:

Co-moving distances to the upper and lower limits of each redshift bin were calculated using the integral expression for the angular diameter distance to a given redshift $D(z)$: 

\begin{equation}
    D(z) = c \int_0^z \frac{dz'}{H(z')}
\end{equation}

\begin{equation}
    H(z') = H_0 \sqrt{\Omega_m (1+z')^3 + \Omega_\Lambda}
\end{equation}

Treating MIRI and NIRCam measured LRDs together; the total COSMOS-Web survey area corresponds to a total solid angle $\Omega = 165 \times 10^{-6}$ sr. A cone with its tip at the observer represents the co-moving volume between the JWST and galaxies at a particular redshift. We thus calculate the co-moving volume $V(z)$ of this cone using the predicted angular diameter distances and the total survey solid angle of view:

\begin{equation}
V(z) = \frac{\Omega}{3} D(z)^3
\end{equation}

To finally determine the total co-moving volume within a given redshift bin, we subtract the co-moving volume at the lower redshift bin limit from the co-moving volume at the upper redshift bin limit. As shown in table \ref{tab:cosmos_flares_comparison}, the COSMOS-Web survey covers a larger volume and includes fewer galaxies than the FLARES simulation across all redshift bins. Galaxy number densities were then calculated for each redshift bin and graphed against stellar mass to obtain galaxy stellar mass functions for COSMOS-Web FLARES galaxies.

Theoretical upper limits corresponding to $\varepsilon>0.2$ and $\varepsilon>1$ for these mass functions were then calculated using Equation \ref{eq:mass_relation}. As galactic halo mass data is not available in the datasets used in this study, a theoretical halo mass function was used and then converted to stellar mass functions corresponding to these specific baryon-to-star conversion efficiencies using Equation \ref{eq:mass_relation}. The front-end Python package of the HMFcalc tool, hmf v2.0.0, was used to generate halo mass functions. The same Planck 2018 \cite{PLANCK} cosmology as before was inputted as a cosmology parameter into this mass function generator to ensure consistency throughout the entire analysis. The 'Behroozi' fitting function was implemented as it is predictive at redshifts $0<z<8$ \cite{Behroozi_2015}. 

\begin{table}[!htb]
\centering
\small 
\begin{tabularx}{\textwidth}{l *{4}{>{\centering\arraybackslash}X}} 
\toprule
\textbf{Redshift Bin} & \textbf{COSMOS-Web Volume (Mpc\(^3\))} & \textbf{COSMOS-Web Galaxy Count} & \textbf{FLARES Volume (Mpc\(^3\))} & \textbf{FLARES Galaxy Count} \\
\midrule
$4.5 < z \leq 5.5$  & $3.33 \times 10^{6}$ & 74  & $1.35 \times 10^{6}$ & 2,144 \\
$5.5 < z \leq 6.5$  & $3.65 \times 10^{6}$ & 182 & $1.35 \times 10^{6}$ & 891 \\
$6.5 < z \leq 7.5$  & $4.03 \times 10^{6}$ & 141 & $1.35 \times 10^{6}$ & 348 \\
$7.5 < z \leq 8.5$  & $4.48 \times 10^{6}$ & 31  & $1.35 \times 10^{6}$ & 121 \\
$8.5 < z \leq 9.5$  & $4.99 \times 10^{6}$ & 5   & $1.35 \times 10^{6}$ & 30 \\
$9.5 < z \leq 10.5$ & $5.59 \times 10^{6}$ & 0   & $1.35 \times 10^{6}$ & 8 \\
\bottomrule
\end{tabularx}
\captionsetup{width=1\linewidth}
\caption{Comparison of COSMOS-Web and FLARES volumes and galaxy counts across different redshift bins.}
\label{tab:cosmos_flares_comparison} 
\end{table}

\subsection{Analysis}

Star-forming history was investigated for the simulated and observed galaxies by graphing stellar mass as a function of redshift and the corresponding age of the Universe. The latter was calculated from redshift data using the astropy.cosmology python package where, again, Planck 2018 results \cite{PLANCK} were used as a baseline cosmology for the sake of consistency. Given the relatively large number of FLARES galaxies concentrated at only six specific redshifts, these data were displayed using box-and-whisker plots, with the COSMOS-Web data points overlaid for comparison with 68th percentile uncertainties. Probability density distributions of galactic stellar masses for FLARES and COSMOS-Web were additionally generated for the entire data set and within individual redshift bins. A Mann-Whitney U test was then performed using the scipy python library to compare the observed and simulated distributions. 

Further, Star Forming Sequence analysis was then performed to investigate the correlation between stellar mass (\(M_{\ast}\)) and star formation rate (SFR) for the entire dataset (over the entire redshift range) for the COSMOS-Web survey and the FLARES simulations. The overall trends in star formation activity were visualized by graphing the relationship between these two variables on a logarithmic scale. The data from both the COSMOS-Web and FLARES datasets were fitted with linear regression lines on a log-log scale. The power-law relationship between SFR and stellar mass for each dataset was then derived using these regression lines to highlight the differences in star formation efficiency.

Additionally, regression lines corresponding to a constant specific star formation rate (sSFR) were graphed to serve as references and illustrate the relationship between the sSFR and stellar mass in both datasets. The COSMOS-Web and FLARES data were then compared and analysed by the amount by which their respective trends correspond to a constant sSFR. The star-forming sequence was thus directly compared across the two datasets using a combined plot that included the principal data points, regression lines, and sSFR reference lines.

\section{Results}
\subsection{Galactic Stellar Mass Function}

\begin{figure}[!htb]
  \centering
  \vspace{0mm} 
  \includegraphics[width=1\linewidth]{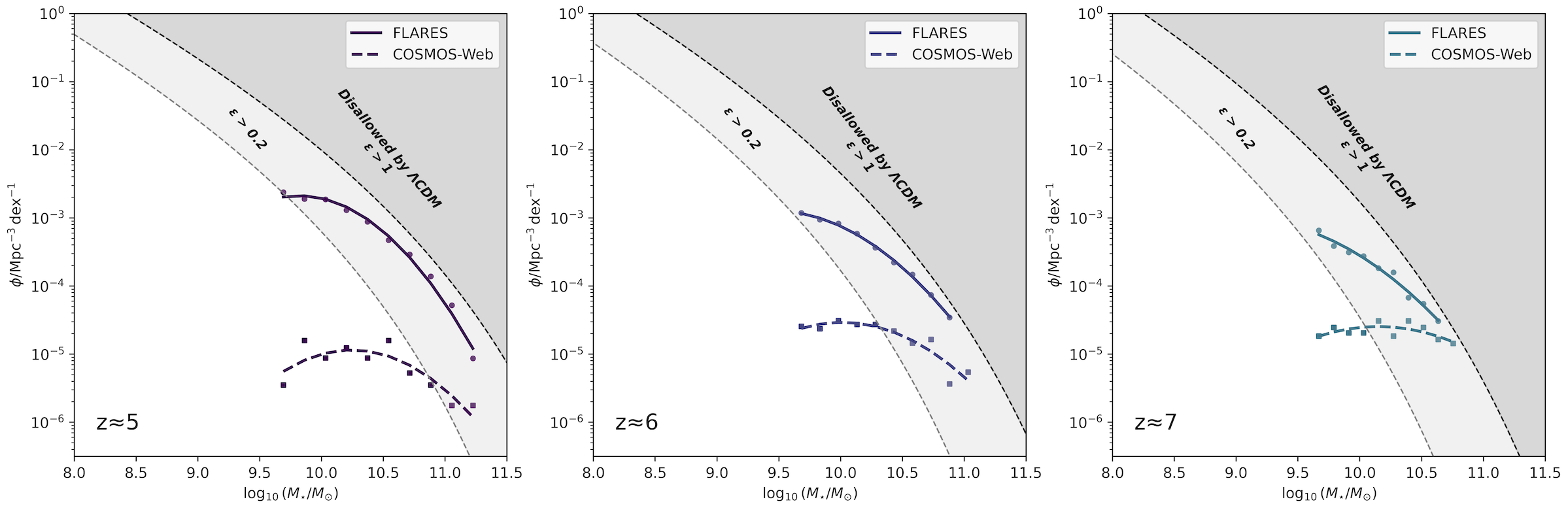} 
  \vspace{-3mm} 
  \captionsetup{width=1\linewidth}
  \caption[Short caption for List of Figures]{Stellar mass functions at $z\approx5$, $z\approx6$, and $z\approx7$ compare galaxy populations derived from the FLARES simulations with those observed in the COSMOS-Web survey. Solid lines represent FLARES data for each redshift in purple, blue, and cyan, respectively, while the corresponding dashed lines denote COSMOS-Web data. Shaded areas indicate stellar mass ranges considered implausible under $\Lambda$CDM cosmology, with light-grey areas showing efficiencies greater than 0.2 and dark-grey areas indicating efficiencies greater than 1, which are disallowed. These figures quantify the number density of galaxies per logarithmic mass interval.}
  \label{fig:GSMF}
\end{figure}

Figure \ref{fig:GSMF} displays the galaxy stellar mass functions for FLARES and COSMOS-Web galaxies at redshifts $z\approx5$, $z\approx6$, and $z\approx7$. The data displayed in Figure \ref{fig:GSMF}, including redshift bin centres and stellar mass bin centres with corresponding galactic number densities, are provided in Appendix A - Table~\ref{tab:appendixA}. The analysis was limited to these redshifts due to insufficient data availability at higher redshifts (COSMOS-Web sample sizes below 100), as evidenced by the data in table \ref{tab:cosmos_flares_comparison}. Squares in Figure \ref{fig:GSMF} represent the binned number densities of observed galaxies in the COSMOS-Web survey, while dots represent those of the FLARES dataset. The minimum and maximum stellar masses in each redshift bin are used to bin the stellar masses of galaxies into ten intervals. The density per dex is then obtained by normalizing the number density of galaxies in each bin by the bin width, which considers the logarithmic nature of the stellar mass axis. 

Solid and dashed lines represent quadratic fits to the FLARES and COSMOS-Web number densities, respectively. The shaded areas represent the upper boundaries where the baryon-to-star conversion efficiency (\( \epsilon \)) exceeds crucial thresholds. The dark-grey areas match \( \epsilon > 1 \), which is physically forbidden as a galaxy's stellar mass would exceed the available baryonic matter in the halo it to form (under the assumption of Planck 2018 concordance cosmology \cite{PLANCK}). The light-grey areas represent baryon-to-star conversion efficiencies of \( \epsilon > 0.2 \) which are higher-than-expected in the local Universe as discussed in Section~\ref{subsec:high_redshift_problem}. These shaded areas thus represent theoretical constraints within the $\Lambda$CDM framework and current models of galaxy formation. 

FLARES galaxy number densities consistently lie above the \( \epsilon > 0.2 \) threshold across all redshifts, although observed LRDs enter this region only at the high stellar mass end. The stellar mass at which observed LRD number densities exceed the \( \epsilon > 0.2 \) threshold also appears to significantly decrease with increasing redshifts. At redshift $z\approx5$, number densities of LRDs with masses above $\log_{10} \left( M_\star / M_\odot \right) \approx 10.9$ exceed the \( \epsilon > 0.2 \) threshold, while at $z\approx7$, the stellar mass limit to cross same the threshold is significantly lower at $\log_{10} \left( M_\star / M_\odot \right) \approx 10.1$. However, simulated and observed galactic number densities consistently lie below the \( \epsilon > 1 \) threshold across all redshifts.

Galactic number densities of FLARES galaxies are consistently several orders of magnitude higher than LRD galaxies of corresponding stellar mass. This discrepancy also appears to be larger by 1-2 orders of magnitude at the low-stellar mass end than at the high-stellar mass end and additionally appears to increase by 1-2 orders of magnitude to lower redshift. A maximum discrepancy of 3 orders of magnitude is found at the lowest stellar mass bin at redshift $z=5$.

Both FLARES and COSMOS-Web galaxy densities appear to exhibit a steep decline at the high-mass end ($\log_{10} \left( M_\star / M_\odot \right) > 10.4$) and a flatter mass function at the low stellar mass tail. This difference is most apparent for the COSMOS-Web galaxies where at $z\approx7$ and $z\approx6$, the number densities are approximately constant in the low mass end and at $z\approx5$, the number LRD densities even increase with stellar mass at the low mass tail of the function. On average, FLARES galactic number densities also appear to decrease with increasing redshift, while COSMOS-Web densities slightly increase with redshift. 

\captionsetup{font=scriptsize} 

\subsection{Star Formation History}

\begin{figure}[!htb] 
  \centering
  \vspace{0mm} 
  \includegraphics[width=1\linewidth]{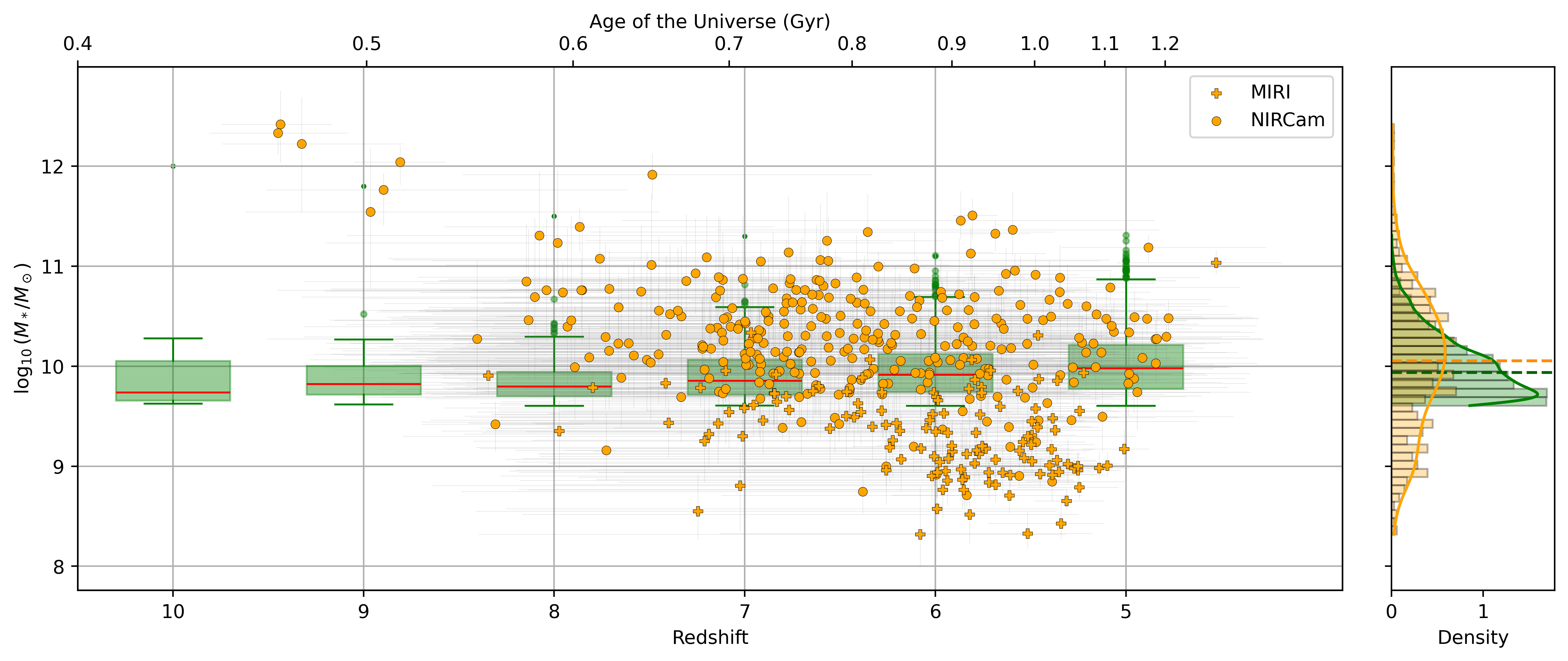} 
  \vspace{-3mm} 
  \caption[Short caption for List of Figures]{Stellar mass distribution of galaxies as a function of redshift and age of the Universe (Gyr). Data points and box-and-whisker plots represent observations from the COSMOS-Web survey and simulations from the FLARES project. Yellow circles (COSMOS-Web) vary in shape based on the JWST instrument, with MIRI data shown as crosses and NIRCam data as squares. Green box-and-whisker plots summarize FLARES simulation data. Stellar masses are displayed on a logarithmic scale. The horizontal dashed lines in the density figure represent median values for both datasets.}
  \label{fig:SIMOBS}
\end{figure}

\begin{figure}[!htb] 
  \centering
  \includegraphics[width=1\linewidth]{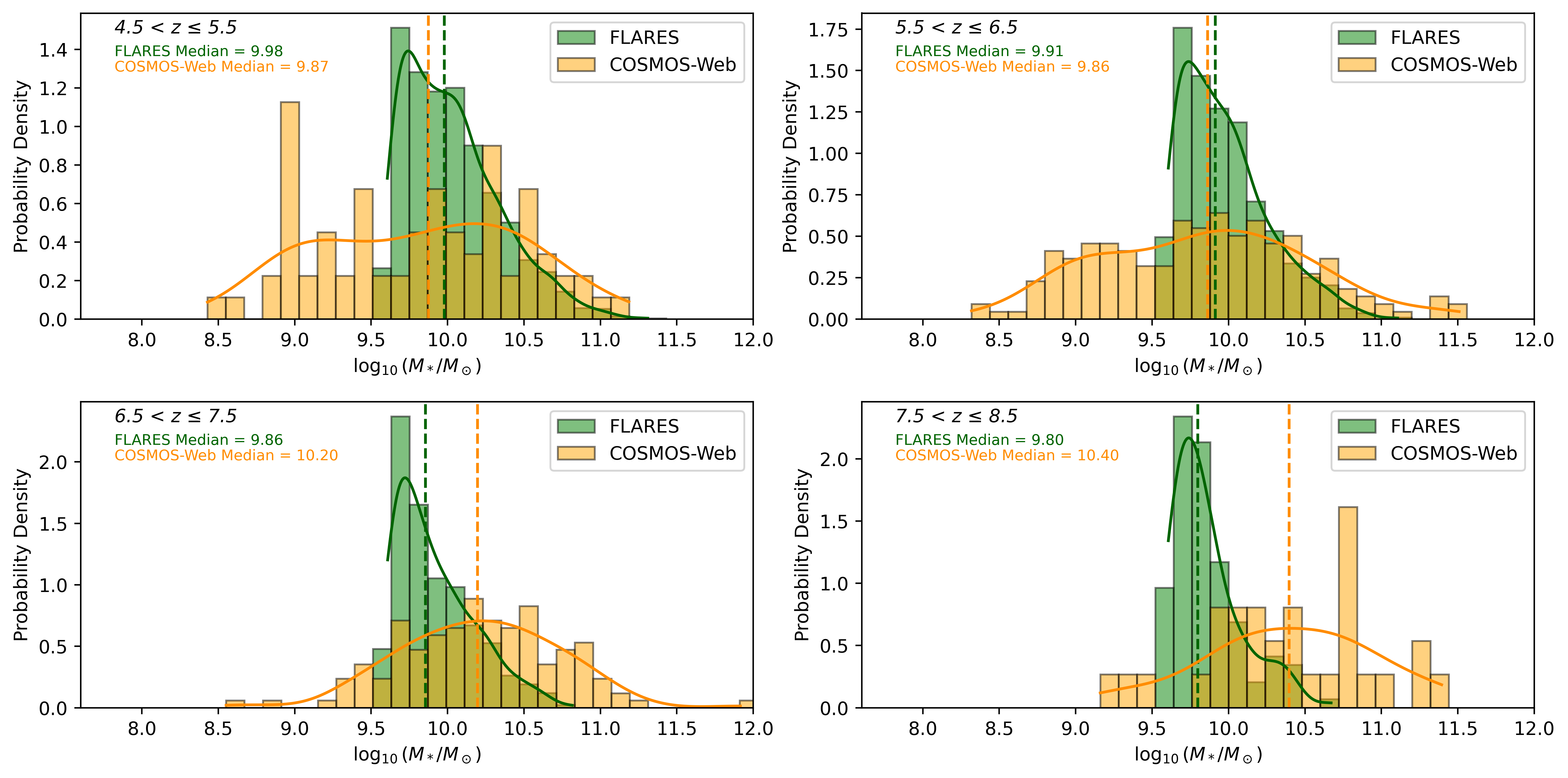} 
  \vspace{-3mm} 
  \caption[Short caption for List of Figures]{Stellar mass distributions across various redshift bins for observed galaxies from the COSMOS-Web survey and simulated galaxies from the FLARES project. The histograms and overlaid KDEs illustrate the probability density of stellar masses within four redshift ranges: $4.5 < z \leq 5.5$, $5.5 < z \leq 6.5$, $6.5 < z \leq 7.5$, and $7.5 < z \leq 8.5$. Green bars and curves represent FLARES simulation data, while yellow bars and curves depict observations from COSMOS-Web. Dashed vertical lines indicate each dataset's median stellar mass values within the respective redshift bins.}
  \label{fig:Horizontal}
\end{figure}

Next, we turn our attention to Figure \ref{fig:SIMOBS}, which displays the stellar mass (in units of $\log_{10} \left( M_\star / M_\odot \right)$) against redshift for FLARES and COSMOS-Web galaxies. On the right of the figure, the stellar mass probability density distribution is displayed for simulated and observed galaxies, along with kernel density estimates and dashed lines representing the median values for each distribution. Both datasets appear to be approximately normally distributed except for the FLARES stellar mass data, which exhibits a sharp cutoff at $\log_{10} \left( M_\star / M_\odot \right) = 9.61$. This cutoff does not represent a physical phenomenon. It is instead a relic of the filtering process that was applied to the FLARES data to generate mock observations (see section ~\ref{subsec:mock}). The density distributions show that the stellar mass medians demonstrate a close alignment with a FLARES median of $\log_{10} \left( M_\star / M_\odot \right) = 9.94$ and a COSMOS-Web median of $\log_{10} \left( M_\star / M_\odot \right) = 10.05$. Figure \ref{fig:Horizontal} shows the density distributions for individual redshift bins, where redshift bins \( 8.5 < z \leq 9.5 \) and \( 9.5 < z \leq 10.5 \) were omitted from analysis due to insufficient data. This figure clearly illustrates that the similarity of means also holds within redshift bins.

The stellar mass distributions across all redshifts as visualised in figure \ref{fig:SIMOBS} were then statistically evaluated using a Mann-Whitney U test. The two datasets have noticeably different sample sizes—3542 mock observations vs 434 real-world observations—and the FLARES data was not normally distributed. Hence, this test was selected. Being non-parametric, the Mann-Whitney U test is appropriate when sample sizes are significantly different and normality cannot be presumed. The test produced a p-value of 0.0546, slightly above the conventional 0.05 significance threshold. This outcome shows that at the 5\% level, there is no statistically significant difference between the distributions.

On the left figure, green box-and-whisker plots show the stellar mass distribution of FLARES galaxies at different redshifts; each box summarises the median values shown by red horizontal lines and the interquartile range. Individual COSMOS-Web galaxies are displayed on top with 68th percentile uncertainties for redshift and stellar mass displayed as grey lines. LRDs captured with the MIRI instrument are represented as orange crosses, while those captured by the NIRCam are represented as orange circles.

Despite significantly large scatter, FLARES and COMOS-Web galaxies show contrasting stellar evolution patterns. FLARES shows a  positive correlation between stellar mass and the Universe's Age. Conversely, COSMOS-Web galaxies demonstrate an approximately constant trend with a small population of exceptionally high stellar mass galaxies at redshifts $z>8$, which have been extensively reported by Akins et al. (2024) \cite{akins2024cosmosweboverabundancephysicalnature}.

\captionsetup{font=scriptsize, width=1\linewidth} 

\captionsetup{font=scriptsize, width=1\linewidth} 

\subsection{Star-Forming Sequence}
\label{subsec:Star-Forming Sequence}

\begin{figure}[!htb] 
  \centering
  \vspace{0mm} 
  \includegraphics[width=1\linewidth]{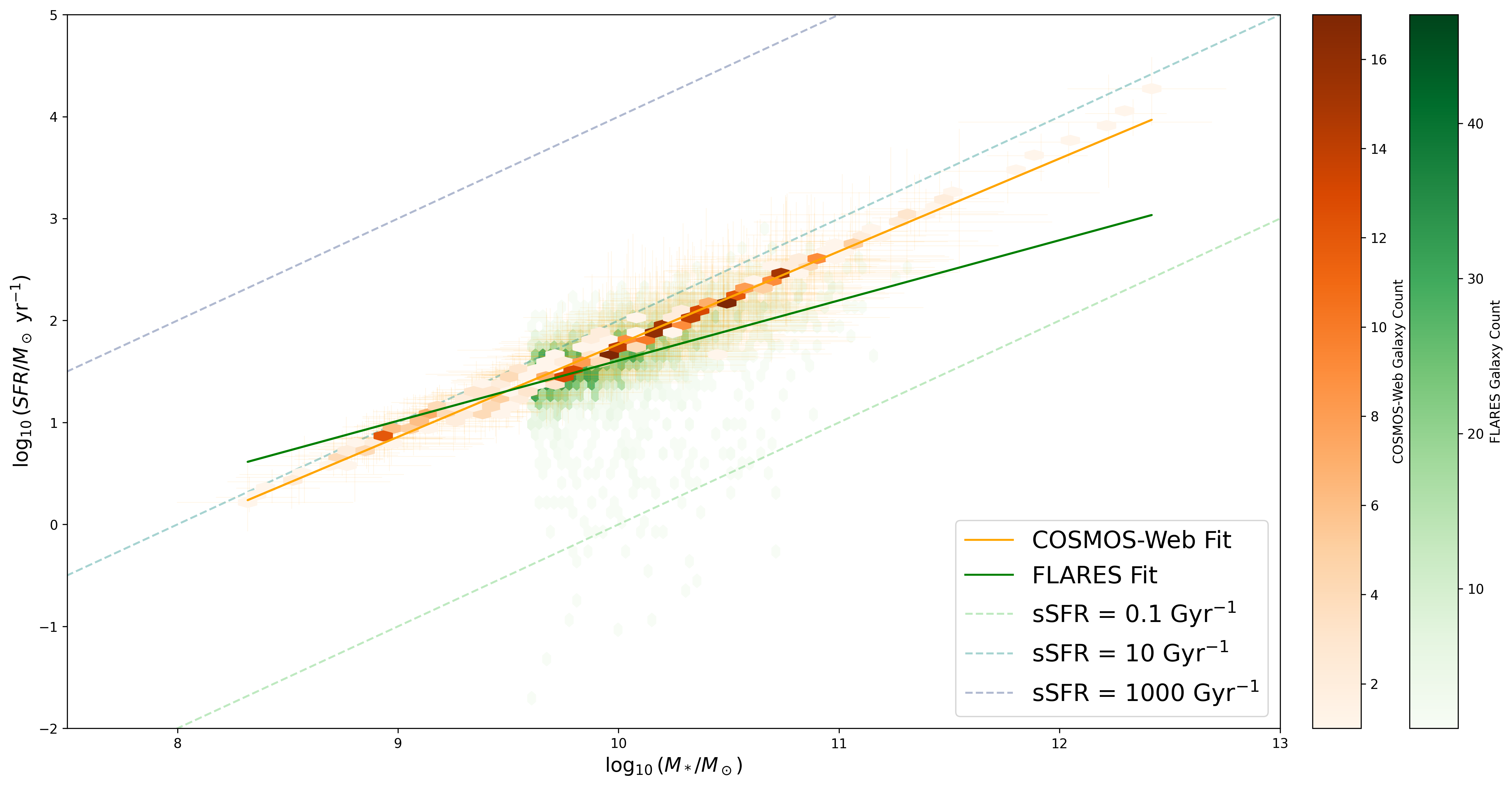} 
  \vspace{-3mm} 
  \caption[Short caption for List of Figures]{Correlation between stellar mass and star formation rate (SFR) for galaxies is displayed using a hexbin method. Colour gradients within hexagonal bins show galaxy density, with darker greens for higher concentrations of galaxies from the FLARES simulation and darker oranges for denser regions of galaxies from the COSMOS-Web survey. Solid lines representing specific star formation rates (sSFR) of 0.1, 10, and \( 1000 \ \text{Gyr}^{-1} \) are superimposed to indicate typical evolutionary tracks.}
  \label{fig:starform}
\end{figure}

Figure \ref{fig:starform} shows the star-forming sequences of FLARES and COSMOS-Web galaxies represented using a green and orange hexbin diagram, respectively. Hexagon shade represents the density of galaxies that fall within a given bin. The COSMOS-Web values are accompanied by 68th percentile uncertainties for SFR and stellar mass, displayed as orange lines. Star-forming sequences corresponding to constant specific star formation rates (star formation rate per unit stellar mass) of \( 0.1 \ \text{Gyr}^{-1} \), \( 10 \ \text{Gyr}^{-1} \) and \( 100 \ \text{Gyr}^{-1} \) are additionally displayed as solid lines. Linear fits to the COMSOS-Web and FLARES data are displayed as solid orange and green lines, respectively. These linear regressions point to different power law relations between SFR and stellar mass for simulated and observed galaxies. The power law relations are given by:

\begin{equation}
\text{COSMOS-Web:} \quad \text{SFR} \approx 4.68 \times 10^{-8} \times M_*^{0.91}
\end{equation}

\begin{equation}
\text{FLARES:} \quad \text{SFR} \approx 5.01 \times 10^{-5} \times M_*^{0.59}
\end{equation}

The COSMOS-Web data shows an approximately linear scaling relation of star formation rate with stellar mass, with an exponent of 0.91. This corresponds to an almost constant specific star formation rate with only a slight decrease with increasing stellar mass given by $\text{sSFR}_{\text{COSMOS-Web}} \approx 4.68 \times 10^{-8} \times M_*^{-0.09}$. 

\begin{figure}[!htb] 
  \centering
  \vspace{0mm} 

  \begin{minipage}{0.5\linewidth} 
    \centering
    \includegraphics[width=\linewidth]{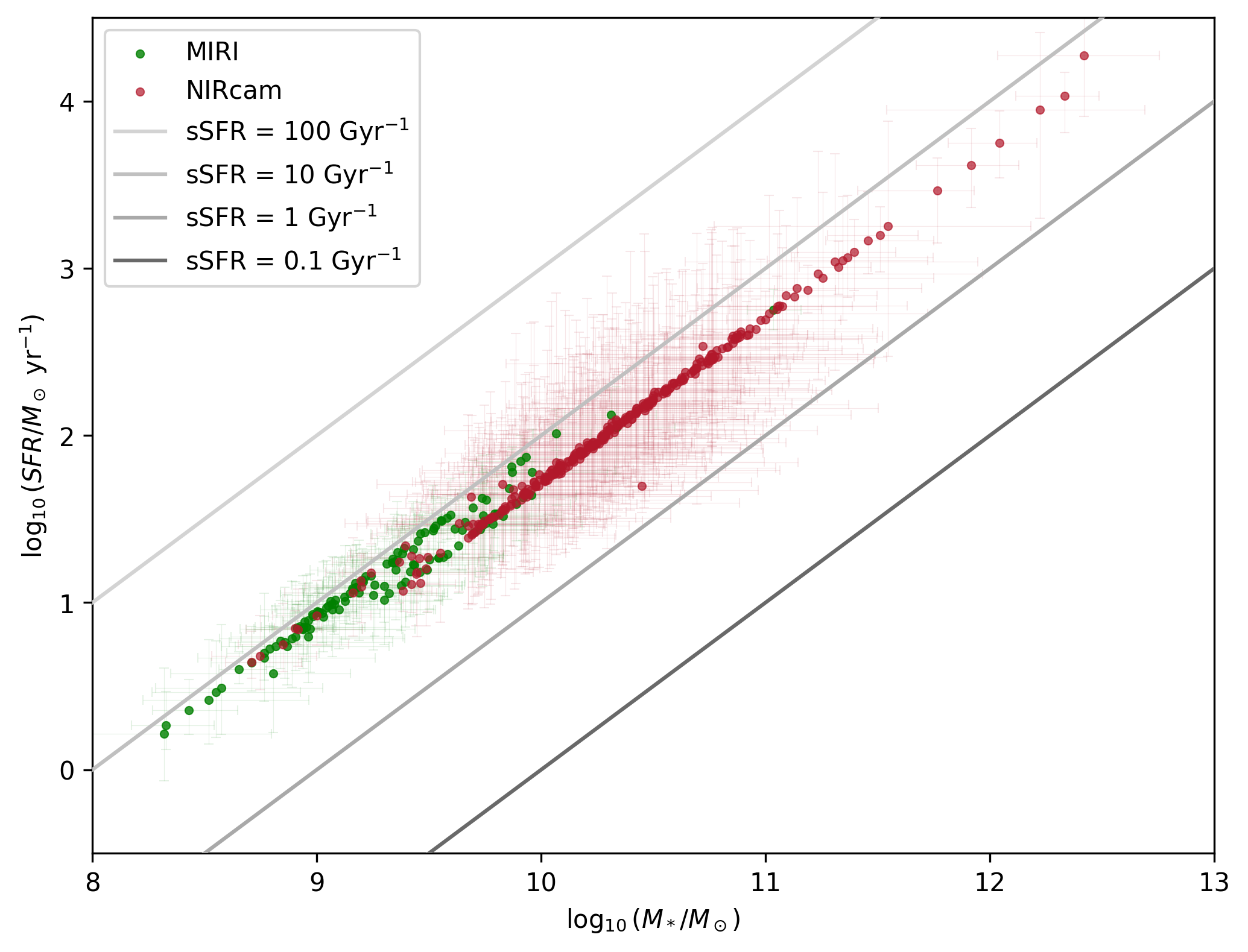} 
    \vspace{-2mm} 
    \captionsetup{width=\linewidth} 
    \caption[Short caption for List of Figures]{Correlation between stellar mass and star formation rate (SFR) for LRDs imaged by the MIRI instrument (green) and the NIRCam instrument (red) in the COSMOS-Web survey. Corresponding lines of best fit are graphed as dashed lines of the corresponding colour.}
    \label{fig:twopop}
  \end{minipage}
  
\end{figure}

In Figure \ref{fig:twopop}, we show an apparent splitting of two separate populations in the COSMOS-Web data. These two populations appear to lie along star-forming sequences corresponding to slightly different constant specific star formation rates. The two populations correspond to galaxies imaged with the MIRI instrument (green) and the NIRCam instrument (red), where the population imaged by MIRI corresponds to a slightly higher constant sSFR. As this population splitting is an unmistakable artefact of JWST measurement, we treat these two populations as a single population in further analysis and assume the absence of any underlying physical phenomenon. This population split can explain the slight deviation from constancy in the specific star formation rate of the entire COSMOS-Web LRD population. The NIRCam-imaged LRDs exhibit a marginally lower sSFR in the high stellar mass range, resulting in a subtle decline in the overall sSFR with increasing stellar mass. Additionally, significant 68th-percentile uncertainties in SFR and stellar mass suggest that the entire COSMOS-Web population can be reasonably treated as corresponding to a constant sSFR.

In contrast, the FLARES power law relation shows a more gradual increase of SFR with increasing stellar mass compared to observations, with an exponent of 0.59.
This indicates a less pronounced response of star formation to increases in stellar mass within the FLARES simulations. Consequently, the FLARES galaxies show a sharper decline in specific star formation rate with increasing stellar mass; $\text{sSFR}_{\text{FLARES}} \approx 5.01 \times 10^{-5} \times M_*^{-0.41}$. These findings are also evident in Figure \ref{fig:starform}, where the orange line parallel to the constant sSFR lines indicates that the specific star formation rate is approximately constant for COSMOS-Web. Conversely, as stellar mass increases, the FLARES data shows a decreasing sSFR, as the green regression line's downward slope shows.

The FLARES dataset shows a significantly higher baseline scaling factor of $5.01 \times 10^{-5}$ than the COSMOS-Web dataset, with a baseline scaling factor of $4.68 \times 10^{-8}$.
These baselines suggest that FLARES consistently predicts higher star formation rates across all stellar masses than observed LRDs by approximately 3 orders of magnitude. The disparity in baselines across the two datasets suggests that FLARES predicts a significantly higher SFR than COSMOS-Web for a given stellar mass. These power-law relations show how star formation processes differ from simulations to observational data. In contrast to the predictions from the FLARES simulations, COSMOS-Web suggests a stronger dependence of SFR on stellar mass, albeit at a lower overall rate.

\section{Discussion}

\subsection{Galaxy Stellar Mass Functions}

Comparative analysis of galaxy stellar mass functions reveals significant tension between simulation and observation. Across all investigated redshifts ($z\approx5$, $z\approx6$, and $z\approx7$), the analysis revealed that the FLARES simulation overestimates number densities by several orders of magnitude compared to observed LRDs.  This significant overprediction of high-redshift galaxy abundance by the FLARES simulation suite aligns with previous findings by Vijayan et al. (2020) \cite{Vijayan_2020}, where simulated galaxies were compared to high-redshift HST observations.

This conclusion is supported by the fact that FLARES galaxies consistently exceed the $\epsilon > 0.2$ threshold, which implies that FLARES additionally overestimates the efficiency of star formation processes or underestimating the impact of feedback mechanisms, such as supernovae and active galactic nuclei (AGN). Feedback processes such as these regulate star formation by heating and expelling gas from galaxies, thereby restricting the rate at which new stars can form \cite{booth2013interactionfeedbackactivegalactic}. A study by Florez et al. (2020) \cite{Florez_2020} shows that when AGN feedback is not accounted for in SED fitting, stellar masses can be overestimated by factors of $\approx100$ in extreme cases. FLARES may thus predict a more significant number of high stellar mass galaxies in the early Universe than are observed, particularly at lower masses where feedback effects are most significant if these processes are not adequately modelled \cite{Florez_2020}. This would additionally explain the increased discrepancy between FLARES number densities at lower redshift, as FLARES galaxies form progressively more stars compared to observations, likely due to the reduced effectiveness of feedback processes.

Despite these discrepancies, FLARES and COSMOS-Web do both demonstrate a significant decrease in the number density of galaxies at the high-mass end. This result agrees with theoretical predictions and prior research (e.g., Kannan et al. (2023) \cite{Kannan_2023}). Moreover, number densities for both simulated and observed galaxies do not test or exceed the $\varepsilon > 1$ threshold imposed by $\Lambda$CDM cosmology.

\subsection{Star Formation History}

Comparing the stellar mass distributions between the simulation and COSMOS-Web data is crucial to assess how well the FLARES simulation replicates stellar attributes of the observed LRDs. The the Mann-Whitney U test p-value of 0.0546 is slightly above the chosen significance level of 5\% and is thus a significant result of this investigation. This p-value suggests a significant overall agreement between simulation and observation, as there is no substantial evidence to support a statistically significant difference between the two distributions.

However, the p-value's proximity to the 5\% significance level raises questions about the possibility of underlying differences that the present analysis may not fully capture. The stellar mass medians are nearly aligned, with $\log_{10}(M_*/M_\odot) = 10.05$ for COSMOS-Web and $\log_{10}(M_*/M_\odot) = 9.94$ for FLARES. This proximity of stellar mass medians is also observed in individual redshift bins, where we observe the closest median stellar mass alignment at redshifts \(5.5 < z \leq 6.5\).

Nevertheless, the near-threshold p-value suggests that there may be significant differences, especially in the distribution tails or in particular galaxy populations. The FLARES simulation's sharp cutoff at lower star masses, which results from the filtering process, introduces a potential bias that could affect the p-value. This filtering step skewed the simulated stellar mass distribution toward higher masses by eliminating galaxies that the JWST could not observe. This thus introduced a source of bias towards higher stellar masses. Without this necessary filtering, the simulation output included more low-mass galaxies, which could have changed the p-value and revealed more significant differences. Improved simulation methods and larger sample sizes may reveal whether or not this apparent agreement masks more significant differences between observed and simulated stellar mass distributions.

No significant upward or downward trend of stellar mass is observed over cosmic time in the COSMOS-Web data. While the overall stellar mass appears relatively constant, there is considerable scatter in the data. This indicated large variability in the star formation history of LRDs, suggesting some galaxies may have formed most of their stellar mass early while others may have continued to grow throughout the redshift range investigated in this study.  On the contrary, a positive trend in the FLARES data suggests an overall increase in stellar mass over cosmic time. This mismatch may again result from the simulation's poor modelling of feedback systems essential to controlling star formation, such as AGN and supernovae. The continuous mass growth shown in FLARES could result from simulated galaxies continuing to form stars for more extended periods of time in the absence of effective feedback.

\subsection{Star Forming Sequence}

Star-forming sequence or 'main sequence' analysis is essential in understanding stellar mass build-up in high-redshift galaxies. It provides valuable information on how star formation efficiencies change as galaxies form stars and a complex array of feedback mechanisms take effect. A significant number of galaxies out to high redshift have been observed to exhibit a tight relation between galactic stellar masses and star formation rates in previous research, although with significant scatter at constant stellar masses due to a complex interplay of factors that increase or suppress star formation across different temporal and spatial scales \cite{Speagle_2014}\cite{Noeske_2007}. Galaxies that have ceased star formation ('passive galaxies') are observed to then "fall off" this main sequence.

The observed LRD star-forming sequence showed a clear two-population split corresponding to imaging instruments. The MIRI-imaged galaxies appear to follow a slightly higher specific star formation, corresponding to a systematically higher star formation rate than NIRCam-imaged galaxies. We assume this to be the result of the differences in wavelength sensitivity between the two instruments. NIRCam is the primary imaging component of the JWST. It records light in the near-infrared spectrum in a wavelength range of 0.6$\mu$m to 5.0 $\mu$m. Conversely, MIRI runs in parallel and records farther into the infrared, picking up mid-infrared wavelengths from 5.6$\mu$m to 25.5 $\mu$m \cite{Rieke_2023}. This allows MIRI to record light from more dust-obscured star-forming regions than NIRCam. Therefore, MIRI's sensitivity to longer wavelengths allows it to record more star formation activity in these usually dust-obscured regions, resulting in systematically higher sSFR measurement.

The star-forming sequence analysis reveals discrepancies between the FLARES simulation and COSMOS-Web data. With an exponent of 0.91, the COSMOS-Web data demonstrate a nearly linear relationship between star formation rate and stellar mass, suggesting an approximately consistent specific star formation rate across the observed range of stellar masses. This result is consistent with other studies of observed high-redshift galaxies, such as those reported by Khusanova et al. (2020)\cite{Khusanova_2021}, who found a nearly constant sSFR at $z>4$. On the other hand, the FLARES simulation shows a weaker star formation rate—stellar mass relation and thus a decreasing specific star formation rate, indicating that the star formation efficiency of simulated galaxies decreases with increasing stellar mass. On the other hand, The FLARES dataset shows a significantly higher baseline scaling factor than the COSMOS-Web observations. This higher baseline could again reflect an underestimation of the effects of feedback mechanisms such as AGN and supernova feedback or other model assumptions. A decreasing sSFR with increasing stellar mass has been observed in the EAGLE simulation at stellar masses above $10^{10} M_{\odot}$. This decrease has been potentially attributed by Furlong et al. (2015) \cite{Furlong_2015} to gas supply exhaustion, where cold gas required for star formation is depleted as galaxies grow more massive. This gradual exhaustion of cold gas would be an independent factor in the decrease in sSFR. 

\subsection{Support of the AGN Scenario}

Comparison of galaxy stellar mass functions, star formation sequences, and star formation histories between observed LRDs and FLARES galaxies consistently point to a possible underestimation of stellar feedback mechanisms such as AGN feedback. AGN activity has been shown to suppress star formation in simulations through the following mechanisms:

\begin{itemize}
    \item Gas heating and expulsion: The heating and expulsion of gas and dust have been associated with AGN activity, where powerful jets, winds and radiation pressure can expel and heat gas in the galaxy \cite{Steinborn_2018}\cite{King_2015}. Cold and dense gas is essential in the collapse of gas into stars as low temperatures reduce thermal pressure, allowing gravity to dominate, while high density enhances gravitational collapse. Therefore, as AGN heat surrounding gas and reduces gas density, it negatively affects overall star formation rates.
    \item Feedback mechanisms: The introduction of AGN activity in simulations has also been shown to limit the effects of the "cooling problem" in galaxy simulations, where gas cooling in dark matter halos is too efficient, leading to runaway star formation \cite{Benson_2003}. Negative feedback from AGN activity then helps prevent this overestimation of stellar activity, such as that observed in the FLARES simulation.
    \item Intracluster medium heating: In addition to gas heating within the galaxy, AGN activity in galaxies within a cluster has also been linked to reduced galaxy formation as the AGN heat the gas between galaxies. This, in turn, reduces the availability of cool and dense gas and prevents new galaxies from forming. This effect could also explain the overestimation of galactic number densities simulated by FLARES.
\end{itemize}

Based on the relatively lower galactic number densities and star formation rates observed in the COSMOS-Web survey, AGN activity may thus play a more critical role than the FLARES simulation suggests. This result casts doubt on the starburst scenario and supports the AGN hypothesis. Simulations with more complex AGN feedback mechanisms could better match simulated outcomes with observational data. Thus, additional research utilising improved hydrodynamic simulations focusing on AGN feedback might determine whether this could reduce the tension between simulation and observation and validate the AGN scenario. Recent studies have found that the spectral energy distributions of LRD samples in the same redshift as investigated here strongly favour the AGN hypothesis \cite{kokorev2024censusphotometricallyselectedlittle}\cite{durodola2024exploringagnfractionsample}. This evidence likely challenges the starburst hypothesis previously proposed, indicating that active galactic nuclei are likely the main generators of the energy output in these galaxies.

\subsection{Sources of Bias}

The methods and data used in this study introduce several sources of bias and uncertainty that must be considered carefully when interpreting results. The first significant source of bias is the generation of mock observations. Here, FLARES galaxies were filtered to exclude galaxies with stellar masses corresponding to absolute UV magnitude of \(M_{\text{UV}} > -20.015\), as per a theoretical mass-to-light ratio. Although essential to ensure the direct comparison between simulated data and observed LRD data, this filtering step introduced a significant bias toward high-stellar mass galaxies. As is evident from the known tight relation between star formation rate and stellar mass discussed in Section \ref{subsec:Star-Forming Sequence}, this bias towards high-stellar mass galaxies introduced a corresponding bias in star formation rate. Thus, the mock observations over-represent galaxies with high star formation rates. Although these sources of bias do not directly impact galactic number densities, they likely affected the results from star-forming sequence analysis. The bias likely resulted in underestimating the increase of SFR with stellar mass in the FLARES data. Consequently, the decline in FLARES sSFR and the discrepancy between baseline SFR might have been exaggerated. The assumption of a constant mass-to-light ratio is also a significant approximation in our analysis. Mass-to-light ratios have been shown to vary over cosmic time, galaxy morphology, metallicity and stellar mass \cite{Santini_2023}\cite{L_pez_Sanjuan_2019}. The mass-to-light ratio used in this investigation was shown to be accurate at high redshift by Grazian et al. \cite{Grazian_2015}. However, a certain level of uncertainty in the mass-to-light ratio due to photometric error and model degeneracies has been acknowledged in their analysis. As such, hidden biases are likely to result from this approximation. 

As detailed in Section \ref{subsec:FLARESsimul}, the FLARES simulation uses an overdensity "zoom-in" strategy to investigate galaxy formation, concentrating on areas of higher-than-average density. This technique adds a sampling bias to the analysis performed in this paper because it over-represents dense, active locations relative to the more expansive, typical cosmic environments represented by the COSMOS-Web survey. Galaxy populations in FLARES are likely denser and more massive than those observed, which causes an overestimation of galaxy number densities and stellar masses. This sampling bias results in a systematic error in the simulation data. As such, this selection of overdense regions could explain some tension between the simulated results and observational data. Specifically, this includes the significant overestimation of galactic number densities by the FLARES simulation compared to observed LRD densities.

Another source of sampling bias inherent to this study is the relatively small sample sizes of both simulated and observed galaxies. These small sample sizes could lead to sampling error. Consequently, it is essential to consider the effects of statistical noise on our results. Compared to spectroscopic redshifts, the reliance on photometric redshifts in the COSMOS-Web data again introduces inherent uncertainties and could significantly affect the star formation histories discussed in this study. Additionally, relatively large errors in stellar mass and star formation rates in the COSMOS-Web data could similarly affect our results.

\section{Conclusion}

The main objective of this investigation was to assess if the observed characteristics of little red dots could be replicated with the FLARES simulation. By comparing star formation history, stellar main sequence and galaxy stellar mass functions between simulated and observed datasets, this study aimed to assess the feasibility of the starburst hypothesis for LRDs, in which the majority of LRD energy output (luminosity) is assumed to be caused by significant stellar activity instead of active galactic nuclei. To make accurate and bias-free comparisons between simulation and observation, mock observations and galaxy stellar mass functions were generated, where particular attention was paid to differences in observed and simulated co-moving volumes. Statistical and visual analysis was then performed, comparing the distributions and correlations between galactic properties for simulated and observed populations. 

The analysis revealed significant areas of tension between the FLARES galaxies and observed LRDs for a range of galactic properties. First, analysis of galactic stellar mass functions revealed that the FLARES simulation consistently overestimates the abundance of galaxies of comparable stellar mass to observed LRDs by several orders of magnitude, particularly at the low mass end. FLARES galaxy number densities also consistently exceed the $\epsilon > 0.2$ threshold, showing tension with theoretical expectations of star formation efficiencies in the local Universe. Secondly, although slightly inversely proportional with stellar mass, FLARES galaxies show a significantly higher baseline specific star formation efficiency than observed LRDs. 

These results possibly imply that the FLARES simulations may not sufficiently consider negative feedback mechanisms such as supernovae and active galactic nuclei (AGN). This conclusion is supported by recent studies investigating spectral energy distributions of large LRD samples \cite{kokorev2024censusphotometricallyselectedlittle}\cite{durodola2024exploringagnfractionsample}. These results imply that the FLARES simulation is mainly unable to model galactic properties of LRDs in the COSMOS-Web survey. Thus, the original assumption that LRD energy outputs are dominated by the effects of stellar activity now appears unsupported, with the AGN hypothesis emerging as a more plausible explanation. However, this is not a definitive result since sources of error, such as sampling bias by the FLARES simulation, inherent errors associated with relatively small observed and simulated sample sizes and uncertainties in galactic properties, must be fully considered.

Future research should investigate the AGN hypothesis further by employing larger LRD samples with galactic properties derived from fitting SED models that strongly support the AGN hypothesis. AGN-fitted LRD properties could be compared directly to hydrodynamic galaxy simulations incorporating more comprehensive AGN feedback mechanisms using a similar methodology in this study. This investigation could alleviate the tensions between simulation and observation discussed in this work. However, the lack of observable X-ray emission from LRDs would still pose a significant challenge to the LRD hypothesis.  Such results would contribute to identifying the primary mechanisms dominating little red dot luminosity.

From an observational perspective, it is critical to increase redshift measurement precision. Although photometric redshifts offer a broad understanding of galactic properties, their intrinsic uncertainties, especially at high redshifts, can incorporate inaccuracies in determining stellar masses and star formation rates. The requirement for follow-up spectroscopic redshift measurements, which provide far higher precision, is developing as the JWST continues its mission. Unlike photometric calculations, spectroscopic measurements are less susceptible to uncertainties caused by complex spectrum energy distributions and dust obscuration, both of which are features of large redshifts (LRDs). By using these more precise measurements, it will be possible to compare observed data with simulated outputs more effectively.

The JWST's multi-wavelength capabilities—particularly those of its NIRCam and MIRI instruments— offer unheard-of glimpses into the hidden star formation zones inside galaxies. The difference in star formation rates found by NIRCam and MIRI, as seen in COSMOS-Web, emphasizes the necessity of more accurately including dust-obscured regions in simulations. This wavelength sensitivity can unveil new aspects of high-redshift galaxies, supplying information that simulations will eventually need to attempt to match. 

Finally, future efforts in the field of high-redshift galaxy formation must concentrate on advancing observational methods as well as models. In order to resolve the conflicts between observed and simulated galaxy populations and ultimately provide more insight into little red dots and other early-universe galaxies.

\section*{Data and Code Availability}

The mock observations and code underlying the methods in this study are accessible, with supporting documentation, at a public GitHub repository: \href{https://github.com/louisarts/Arts24LRD}{github.com/louisarts/Arts24LRD}. The raw FLARES simulation data were sourced from the FLARES GitHub repository and are made publicly available by FLARES collaborators in hdf5 format at \href{https://flaresimulations.github.io/#data}{flaresimulations.github.io/data}. The raw LRD data underlying this article are made publicly available by Akins et al. (2024) \cite{akins2024cosmosweboverabundancephysicalnature} in CSV format at a public GitHub repository: \href{https://github.com/hollisakins/akins24_cw}{github.com/hollisakins/akins24\_cw}. 

\begingroup
\small
\bibliography{references}

\endgroup
\newpage

\section*{Appendix A}

\begin{table}[htbp]
    \centering
    \resizebox{\textwidth}{!}{%
    \begin{tabular}{c | c c c}
        \textbf{Redshift} & \textbf{Stellar Mass ($\log_{10}(M_\star/M_\odot)$)} & \textbf{FLARES Number Density (Mpc$^{-3}$ dex$^{-1}$ $\times 10^{-5}$)} & \textbf{COSMOS-Web Number Density (Mpc$^{-3}$ dex$^{-1}$ $\times 10^{-5}$)} \\
        \hline
          &   &   &   \\
         & 9.691 & 239.133 & 0.210 \\
         & 9.862 & 190.091 & 0.945 \\
         & 10.032 & 186.619 & 0.525 \\
         & 10.202 & 130.199 & 0.735 \\
         & 10.373 & 87.667 & 0.525 \\
        5 & 10.543 & 47.306 & 0.945 \\
         & 10.713 & 29.078 & 0.315 \\
         & 10.884 & 13.888 & 0.210 \\
         & 11.054 & 5.208 & 0.105 \\
         & 11.224 & 0.868 & 0.105 \\
          &   &   &   \\
          \hline
          &   &   &   \\
         & 9.681 & 118.283 & 1.867 \\
         & 9.831 & 94.134 & 1.734 \\
         & 9.981 & 82.798 & 2.267 \\
         & 10.131 & 59.142 & 2.001 \\
         & 10.281 & 36.471 & 2.001 \\
        6 & 10.431 & 22.178 & 1.600 \\
         & 10.581 & 14.785 & 1.067 \\
         & 10.731 & 7.393 & 1.200 \\
         & 10.881 & 3.450 & 0.267 \\
         & 11.031 & - & 0.400 \\
          &   &   &   \\
          \hline
          &   &   &   \\
         & 9.669 & 65.528 & 1.664 \\
         & 9.789 & 38.582 & 2.219 \\
         & 9.910 & 31.233 & 1.849 \\
         & 10.031 & 27.559 & 1.849 \\
         & 10.151 & 18.372 & 2.774 \\
        7 & 10.272 & 15.923 & 1.664 \\
         & 10.393 & 6.737 & 2.774 \\
         & 10.514 & 5.512 & 2.219 \\
         & 10.634 & 3.062 & 1.479 \\
         & 10.755 & - & 1.294 \\
          &   &   &   \\
         \hline
          &   &   &   \\
    \end{tabular}%
    }
    \caption{Galactic number densities as a function of stellar mass, presented for three distinct redshift bins: $z=5$, $z=6$, and $z=7$.}
    \label{tab:appendixA}
\end{table}

\end{document}